\documentclass[portrait,iop,revtex4]{emulateapj}

\def\bd28{BD+28$^{\circ}$\,4211}
\newcommand{\teff}{$T_{\rm eff}$}
\newcommand{\gta}{\lower 0.5ex\hbox{$ \buildrel>\over\sim\ $}}
\newcommand{\lta}{\lower 0.5ex\hbox{$ \buildrel<\over\sim\ $}}
\newcommand{\nhe} {$N$({\rm He})/$N$({\rm H})}
\newcommand{\msun}{$M_{\rm \odot}$}

\newcommand{\Bd}{BD$+$28$\arcdeg$4211}
\usepackage{natbib}
\usepackage{graphicx}
\usepackage{url}
\usepackage{amsmath}

\begin{document}

\title{A NLTE Analysis of the Hot Subdwarf O Star BD$+$28$\arcdeg$4211.\\
I. The UV Spectrum}

\author{M. Latour\altaffilmark{1}, G. Fontaine\altaffilmark{1},
 P. Chayer\altaffilmark{2}, and P. Brassard\altaffilmark{1}}

\altaffiltext{1}{D\'epartement de Physique, Universit\'e
  de Montr\'eal, Succ. Centre-Ville, C.P. 6128, Montr\'eal, QC H3C 3J7,
  Canada; marilyn@astro.umontreal.ca} 
\altaffiltext{2}{Space Telescope Science Institute, 3700 San Martin
  Drive, Baltimore, MD 21218}
  
\begin{abstract}
  
We present a detailed analysis of the UV spectrum of the calibration
star \Bd~using high-quality spectra obtained with the $HST$ and $FUSE$
satellites. To this aim, we compare quantitatively the observed data
with model spectra obtained from state-of-the-art NLTE metal
line-blanketed model atmospheres and synthetic spectra calculated with
TLUSTY and SYNSPEC. We thus determine in a self-consistent way the
abundances of eleven elements with well-defined lines in the UV, namely
those of C, N, O, F, Mg, Si, P, S, Ar, Fe, and Ni. The derived
abundances range from about solar to 1/10 solar. We find that the
overall quality of the derived spectral fits is very satisfying. Our
spectral analysis can be used to constrain rather tigthly the effective
temperature of \Bd~to a value of \teff = $82,000 \pm 5000$ K. We also
estimate conservatively that its surface gravity falls in the range log $g$ =
6.2$_{-0.1}^{+0.3}$. Assuming that the $Hipparcos$ measurement for
\Bd~is fully reliable and that our model atmospheres are reasonably
realistic, we can reconcile our spectroscopic constraints with the
available parallax measurement only if the mass of \Bd~is significantly
less than the canonical value of 0.5 \msun~for a representative post-EHB
star.

\end{abstract}


\section{INTRODUCTION}

\Bd~is a hot subwarf O (sdO) star whose brightness, high effective
temperature, and relatively simple spectrum have made it a standard
star in the optical domain as well as a calibration star for UV space
missions such as $IUE$, $HST$, and $FUSE$.  Its status of standard
star implies that some of its observationnal properties are very well
known. For example, high precision $UBVRI$ magnitudes have been
presented by \citet{land07}. In addition, its parallax mesurement in the
$Hipparcos$ catalogue places the star at 92$_{-11}^{+14}$ pc. While
studying \Bd~as a spectrophotometric standard, \citet{mas90} found it to
have a faint red companion at a separation of 2.8$\arcsec$.  Moreover,
\Bd~has been extensively observed in the UV range by the missions
mentionned above and there are highly valuable data available on that
star. Ultraviolet spectra are precious tools for studying the chemical
composition via the numerous metallic lines present in this wavelength
range.  It could be thought that with this privileged status, \Bd~would
have been thoroughly studied and its physical parameters would be
accurately known, but this is not exactly the case. In this connection,
it has to be mentionned that because of its high effective temperature
(around 80,000 K), the local thermodynamic equilibrium (LTE)
approximation is inappropriate for model atmospheres intended to
represent the star.  Instead, the more sophisticated and realistic
approach of non-local thermodynamic equilibrium (NLTE) has to be used.
Given the physical and technical difficulties associated with that
approach, however, more efforts remain to be made along that avenue in
order to characterize better the atmosphere of this star.

The first determination of the effective temperature of \Bd, using
quantitative spectroscopy, has been made by \citet{nap93} who estimated
it to be around 82,000 K. He determined this value by comparing the
Balmer lines of the star with those of NTLE model atmospheres with a
variable H/He ratio (but with no metals) for different effective
temperatures and surface gravities. This method is now largely used and
its ability to give reliable fundamental parameters (\teff, log $g$, and
sometimes also \nhe) for white dwarfs and hot subwarf B stars has
already been demonstrated \citep{ber92,saf94}.  However, this method is 
rather tricky in the case of \Bd~because, like others sdO stars and
white dwarfs that have high temperatures, the model spectra used
generally suffer from the so-called Balmer line problem. That is, the
observed lines cannot be simultaneously matched with a unique model
spectrum. In other words, each line needs a model of different effective
temperature in order to be well reproduced. Usually, the lowest lines in
the series (like H$\alpha$ and H$\beta$) need a lower temperature,
while the highest ones are better reproduced at higher temperatures. In
the case of \Bd, to give the ``extreme'' values, H$\alpha$ was best
reproduced at \teff $\simeq$ 50,000 K and H$\epsilon$ at around 
85,000 K \citep{nap93}.  The author also found a log $g$ value 
of 6.2 and a solar helium abundance to be appropriate values for \Bd.  
His best estimate of the effective temperature of $\sim$ 82,000 K was
subsequently confirmed by \citet{dre93} who checked that parts of the
$IUE$ UV spectrum of \Bd~, showing Fe~\textsc{vi} and Fe~\textsc{vii}
lines, are properly reproduced at around 82,000 K. This time, the NLTE
model atmospheres they used included metals, namely carbon, oxygen,
nitrogen, and iron group elements that were grouped together into 6
model atoms (one for each ionization stages between \textsc{iii} and
\textsc{viii}).  With similar models, and on the basis of the same $IUE$
spectrum, \citet{haas96} estimated the abundance of iron to be about ten times
subsolar while nickel was found to be nearly solar. According to them,
oxygen and nitrogen also have abundances near the solar value. The
improved UV spectra taken with $STIS$ onboard the HST allowed to derived
a solar abundance of manganese, while a sole line of chromium indicated
an abundance between two and four times the solar value \citep{ram03}. 

The inclusion of metallic elements in NLTE model atmospheres, though
costly in terms of computation time and complexity, allows not only for
more realistic models, but also permits to solve, at least in part, the
Balmer line problem \citep{wer96}. We will address this issue in more
details in the second paper of this serie (Latour et al. 2013, in
preparation). The optical spectrum of \Bd~is rather featureless in
comparison to its rich UV spectrum. Except for the Balmer and
He~\textsc{ii} lines, nothing else is seen at medium resolution in the
optical range. Because of this, \Bd~was chosen to be part of an
investigation of diffuse interstellar bands in OB stars given its
uncomplicated spectrum. However, the high resolution HIRES spectra of
the star obtained for this investigation at the Keck I telescope show a
lot of narrow absorption lines as well as a handful of emission lines
\citep{her99}. The sharpness of the absorption lines allowed to set an
upper limit on the star's rotational velocity, $v$ sin $i$ $\lta$ 4 km
s$^{-1}$, which is quite slow. 

As for the high-quality $FUSE$ spectrum of \Bd, it has only been used to
study interstellar abundances in the line of sight of the star \citep{son02}. 
To our knowledge, this data set has not been exploited so far to better
characterize the star. Although we have a general idea of its
atmospheric chemical composition, no comprehensive or systematic studies
were made on that star. With high-quality data available from $HST$ and
$FUSE$ in particular, we felt we should exploit them in order to reexamine
the chemical composition of \Bd~and also try to constrain better its
effective temperature and surface gravity by studying the ionization
equilibrium of some metallic elements. Getting a portrait of the
chemical composition of \Bd~should also be a first step in studying the
optical spectrum of the star, with appropriate model atmospheres, for
the Balmer line problem.   

In the second section of this paper we describe our model atmospheres. 
This is followed by a short description of the observational material we
used in Section 3, and of our abundance analysis in Section 4. We then
discuss our attempts at constraining the effective temperature and the
surface gravity by using the ionization equilibrium of metals in the UV
range as well as the parallax distance in Section 5. Finally, we
present a discussion and conclusion in Section 6. 

\section{MODEL ATMOSPHERES}

\subsection{Characteristics of our Model Atmospheres}

We have developed the capacity to compute large grids of NLTE metal
line-blanketed model atmospheres over reasonable timescales (days to
weeks) with our parallel versions of TLUSTY and SYNSPEC that run on a
dedicated cluster of computers (currently containing 320 processors). 
Our setup is described in more details in \citet{lat11} and has not
changed since, apart from the increase in the number of processors we
have available.  Our final fully-blanketed model 
atmospheres for \Bd~include the following ions (besides those of H and
He) : C~\textsc{ii} to C~\textsc{v}, N~\textsc{ii} to N~\textsc{vi},
O~\textsc{ii} to O~\textsc{vii}, Si~\textsc{iii} to Si~\textsc{v},
P~\textsc{iv} to P~\textsc{vi}, S~\textsc{iii} to S~\textsc{vii},
Fe~\textsc{iv} to Fe~\textsc{viii} and Ni~\textsc{iii} to Ni~\textsc{vii}. 
The highest ionization stage of each element is taken as a one-level
atom. More information on the model atoms we used can be found on
TLUSTY's web site\footnote{http://nova.astro.umd.edu/Tlusty2002/tlusty-frames-data.html}
and in Lanz \& Hubeny (\citeyear{lanz03}, \citeyear{lanz07}). 
Since our thorough examination of \Bd's UV spectrum revealed also lines of
argon, magnesium and fluorine, we needed to construct additional model
atoms in order to study the abundance of these extra elements. This was
done with the MODION
program\footnote{http://idlastro.gsfc.nasa.gov/ftp/contrib/varosi/modion/README}
which uses the TOPBASE data \citep{lanz96}. This program allows the
user to choose the explicit energy levels and also build superlevels 
that are included in the model atom. We thus constructed in this way
model atons for the following ions : F~\textsc{iii} with 9 levels and 5
superlevels, 
F~\textsc{iv} with 11 levels and 5 superlevels, F~\textsc{v} with 19
levels and 3 superlevels, F~\textsc{vi} with 12 levels and 5
superlevels, Mg~\textsc{iii} with 37 levels and 3 superlevels,
Mg~\textsc{iv} with 29 levels and 5 superlevels, Mg~\textsc{v} with 18
levels and 2 superlevels, Ar~\textsc{iv} with 39 levels, Ar~\textsc{v}
with 25 levels, Ar~\textsc{vi} with 20 levels, and Ar~\textsc{vii} with
18 levels. All transitions (bound-bound and bound-free) between these
levels are thus considered when the ions are included in a model
atmosphere.  

Since \citet{wer96} underlined the importance of using Stark profiles
for CNO lines when modeling atmospheres of hot stars such as \Bd, we
inspected our different model atoms to check what kinds of profiles were
used. We found that the strongest transitions (often resonance lines) of
each ion are treated with Stark profiles while the weaker ones are
represented by Doppler profiles. We then examined a synthetic spectrum
that could represent \Bd~and identified its most prominent lines and
made sure they were described with a Stark profile in the corresponding
atomic model. This way we added a classic Stark profile to a few more
lines of some elements, namely C~\textsc{iv}, N~\textsc{iv},
O~\textsc{iv} and O~\textsc{v}, Si~\textsc{iv} and P~\textsc{v}. A
striking effect that \citet{wer96} noticed when including Stark profiles
was the disappearance of a high-temperature bump around log $m$ of $-$3
g cm$^{-2}$ which was present when using Doppler profiles only (see his
Figure 1). Since this bump is not seen either in our models (see our
Figure 1), we believe our atomic data are appropriate for the study of
\Bd~or other hot stars. Note that, even without modifying the original
model atoms used by Lanz \& Hubeny (\citeyear{lanz03},
\citeyear{lanz07}), our temperature structures do not present this bump
(see Figure 4 of \citeauthor{lat11} \citeyear{lat11}). 

The inclusion of our metallic elements must be done ``step by step'' if
we want to assure the convergence of our models. Too drastic changes in
the physical parameters of the model atmosphere used as input and the
one we want to compute will prevent the latter from converging. 
Thus, when constructing a grid of these line-blanketed models, we end up
with a number of ``subgrids'' including only some of the elements
mentionned above. In the case of \Bd, we built five ``subgrids'' in
order to end up with our final fully-blanketed one. For example, we have
a grid including only C, N, and O in solar abundances, from which comes
one of the models plotted in Figure 1.  

\begin{figure}[t]
\epsscale{1.1}
\plotone{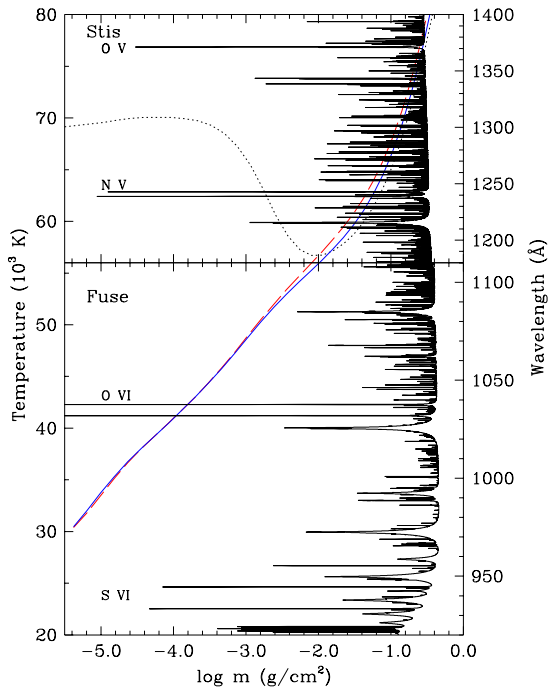}
\caption{Temperature stratification and monochromatic
optical  depth $\tau_{\nu}$ = 2/3 as functions of depth, where {\it m}
is the column  density, for NLTE models defined by ${\it T}_{\rm eff}$ =
82,000 K, log $g$ = 6.2,  and log {\it N}(He)/{\it N}(H) = $-$1.0. 
The temperature structure is shown for models including H and He only (dotted
curve), H, He, and CNO in solar abundances (solid curve), and H, He, and
CNOSiPSFe in solar abundances (dashed curve). The $\tau_{\nu}$ = 2/3 curve
is from the latter model and shows part of the FUSE and STIS wavelenght 
range in the UV.}
\end{figure}

In Figure 1, we show the temperature stratification for models with
\teff =  82,000 K, log $g$ = 6.2, and having a solar helium abundance
culled from three of our grids. These estimates of the atmospheric
parameters come from the work of \citet{nap93}. The first model is a
``classical'' pure H+He NLTE model that shows the well known outwardly
rise of temperature near the surface (dotted curve). The second one
includes C, N, and O (solid curve), while the third one includes all the
elements of our final model (see above) besides nickel (dashed
curve). At this point, all our elements have a solar 
abundance \citep{gre98}. Though we plotted only three models in the
figure, we examined the ones (having the same parameters) from our other
grids and concluded that adding S, P and Si to the C, N, O, only induces
a minor drop of the temperature in the outer layers ( log $m < -$2
). When we add nickel to models already including C, N, O, Si, P, S, and
Fe, the changes in the temperature structure are unnoticeable on a graph
like Figure 1. The cooling of the outer layers is thus mostly done by
the inclusion of C, N, O elements ; adding more metals does not cool
anymore the surface. As for the inner layers, their heating comes 
essentially from the presence of C, N, O, as well as Fe that was
included afterward in the third model depicted in the figure. 

It is easy to see from the temperature stratification that metallic
elements, though they are not the dominant ones in the atmosphere, have
an important effect on the thermodynamic structure at the surface of the
star. Via their important opacity, they block a non-negligible part of
the flux in the UV range, thus rising the continuum in the optical
range. The presence of metals also causes a heating of the inner layers
of the atmosphere, while it cools the outer layers. These changes in the
atmospheric structure influence the emergent spectrum of the star, and
thus the Balmer lines themselves. This is why their presence is an 
essential ingredient in the solution of the Balmer line problem
\citep{wer96}.   

The other curve featured in Figure 1 shows the optical depth $\tau_{\nu}
= 2/3$ as a function of the column density $m$. It allows us to locate
where the continuum and different lines are formed. The most opaque features,
formed very near the surface, are the core of the resonance doublets of
S~\textsc{vi}, O~\textsc{vi} and N~\textsc{v} as well as the
O~\textsc{v} line at 1371 \AA. As for the broadest lines, they are
either hydrogen or helium~\textsc{ii} ones. Depending on where a line is
formed in the atmosphere, it might be affected differently by a change
of the temperature structure. 

In a model atmosphere with fundamental parameters like those of \Bd,
some of the atomic species have a sole ion which dominates the
atmosphere while the other ionization degrees have populations that are
lower by orders of magnitude. This is the case for carbon, silicium and
phosphorus, where ions in a noble gas configuration (C~\textsc{v},
Si~\textsc{v} and P~\textsc{vi}) are the ones that contribute the most
to the thermodynamic structure of the atmosphere.  The population of the
other elements is dominated by ions of different ionization degrees
depending on the depth in the atmosphere.  We show the ionization
equilibria of nitrogen and iron in Figure 2 for models including all the
elements mentionned at the beginning of this section and having an
effective temperature of \teff = 82,000 K (solid lines) and \teff =
92,000 K (dashed lines), a surface garvity log $g$ = 6.2, and a solar
helium content. The temperature is also shown in black 
for each model. Though the equilibrium changes as function of depth,
N~\textsc{v} is dominant in the line forming region (around log $m$ =
$-$1) while Fe~\textsc{vi} and \textsc{vii} are in comparable proportions
in this region for the model at 82,000 K. However, when the
temperature is raised to 92,000 K, Fe~\textsc{vii} become more
dominant. 

\begin{figure*}[t]
\epsscale{1.1}
\plottwo{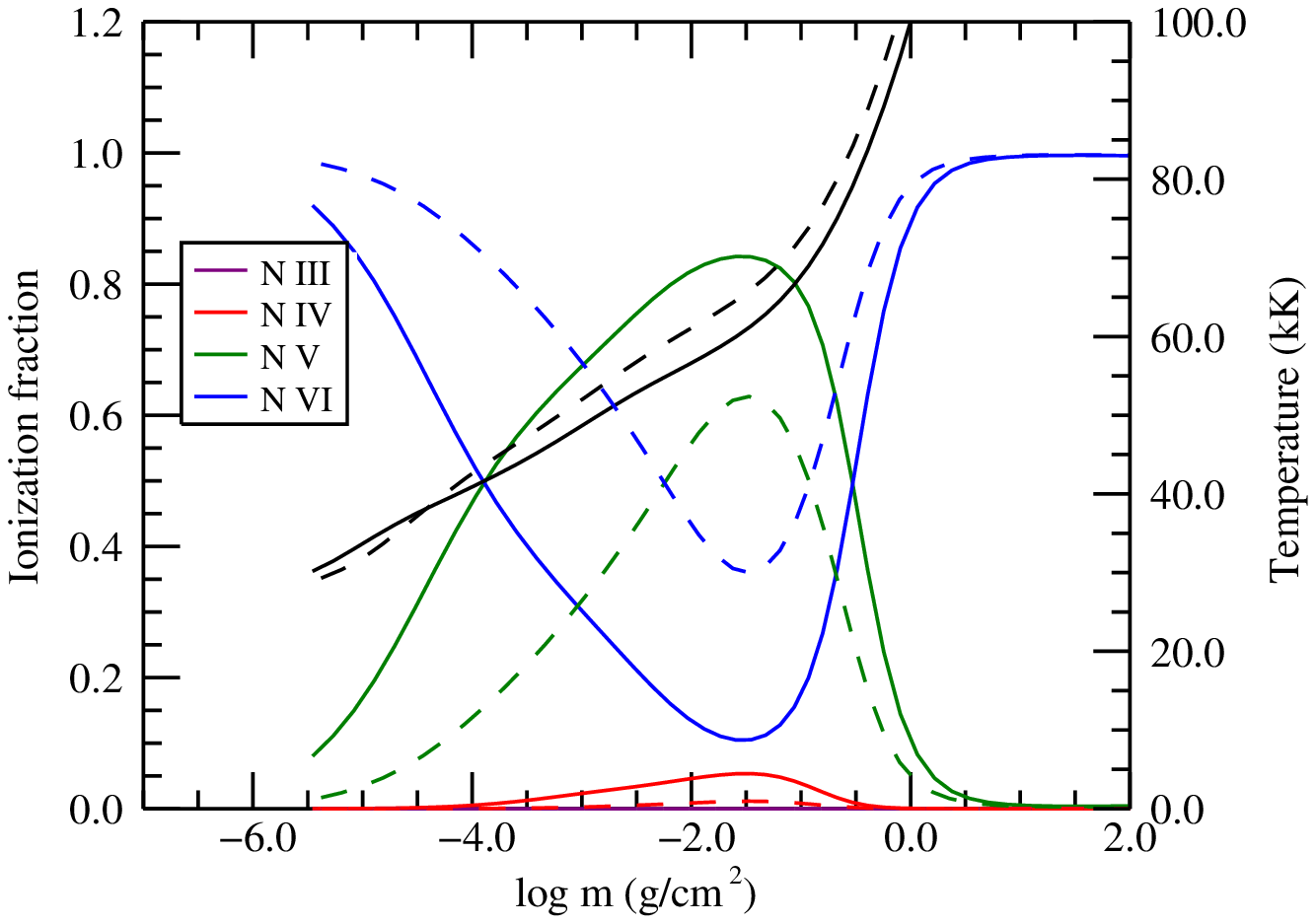}{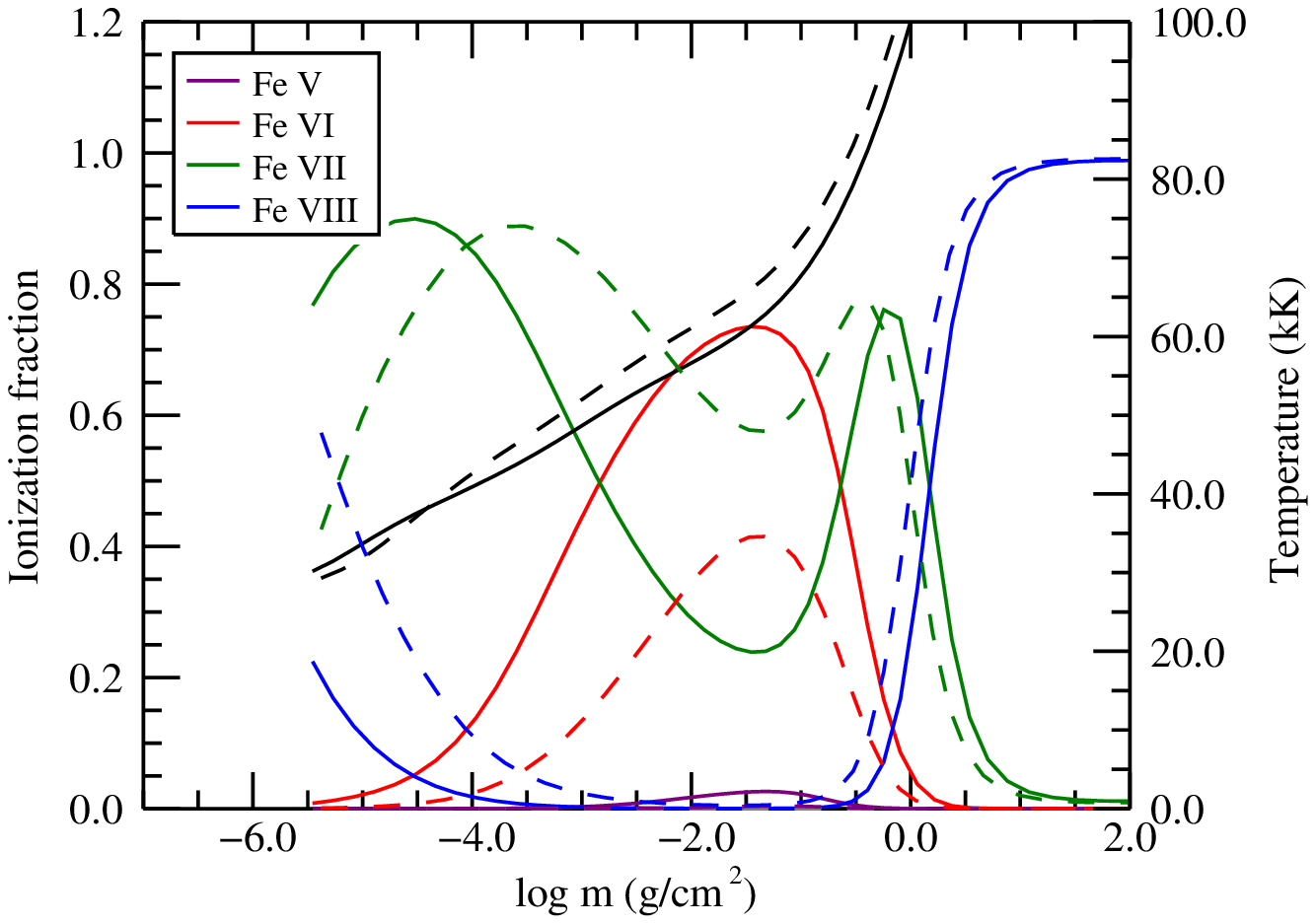}
\caption{ a) Ionization fractions of nitrogen as functions of
depth in models with \teff~ = 82,000 K (solid lines) and 92,000 K
(dashed lines), log $g$ = 6.2, and log \nhe~= $-$1.0. The black curves
show the temperature profiles of the two models. 
b) Similar to Fig. 2a, but this time showing the
ionization structure of iron.  }

\end{figure*}

\subsection{A Word on the Inclusion of Metallic Elements}

By self-consistently including each element we want to study in our
models, we allow it to influence the thermodynamical structure of the
atmosphere. This is crucial if we want to get the right ionization
equilibria and populations for our different ions. \citet{dre93} show
that large discrepancies can arise in the strength of iron lines in a
synthetic spectrum, depending on whether iron is included in a NLTE and
self-consistent way or not. The inconsistent approach uses LTE statistics
to obtain iron populations and ignore the back-reaction of the element
on the atmospheric structure. The discrepancies increase between the two
methods as the NLTE effects increase (as well as the abundance of the element
in question), which means in our case, when the effective temperature
increases. For \Bd, it is essential that the elements we want to fit be
consistently included in the models. 

The TLUSTY package allows to add new elements afterward in the synthetic
spectrum computed by SYNSPEC. Even though it is an inconsistent
approach, we tried it just to take a look at some lines of additional
elements we did not plan to include in our models because of their
expected low abundances, the lack of model atoms, as well as the limit
on the number of atomic levels and transitions the code can handle
without becoming unstable. Even though for some atomic species we could
easily make model atoms with the TOPBASE data, it is not the case for
the iron-peak elements. They are not present in TOPBASE (except for
iron) and their data (for iron and nickel) come from the Kurucz atomic
data sets. Building model atoms for other elements of the iron peak
would be a far more complex task than what was done for magnesium,
fluorine, and argon. That being said, we added a solar amount of manganese, 
cobalt, and chromium in SYNSPEC to an already line-blanketed model 
atmosphere computed without these elements. When the resulting spectrum
was compared with observations, it immediately appeared that there was a
problem with the ionization equilibrium. Both Mn and Cr show lines of two
differents ions (\textsc{v} and \textsc{vi}); lines corresponding to the higher
ionization degree were well reproduced by our model while the ones from
the lower degree were not correctly matched.  In order to illustrate
this, we included nickel in our synthetic spectrum  with the determined
abundance of this element (see section 4.1) and compared the result with
our final model, which included nickel in the model atmosphere. We show
in Figure 3 two regions of the STIS spectrum featuring some
Ni~\textsc{v} and~\textsc{vi} lines. The two upper panels show the
comparison between the observed spectrum of \Bd~and the synthetic one
obtained by adding nickel (log $N$(Ni)/$N$(H) = $-$6.0) in SYNSPEC
only. The lower ones show the result of adding the same amount of nickel
this time directly in the model atmosphere. In the case of nickel,
Ni~\textsc{vi} lines are definitely not strong enough in the upper
panels, which is consistent with the observations of \citet{lanz03} that
NTLE effects favor higher ionization degree due to the strong radiation
field coming from the hotter and deeper layers that causes
overionization. In the other hand, the Ni~\textsc{v} line shows in the
second portion of the spectrum, as well as the few other we examined,
remain quite unaffected by the difference in the inclusion method.  

\begin{figure}[b]
\includegraphics[scale=.50,angle=270]{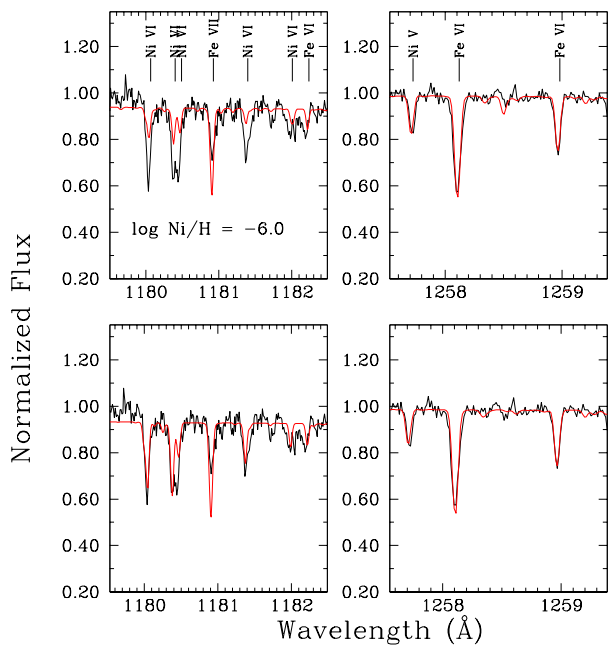}
\caption{ Comparison between portions of the observed STIS spectrum 
of \Bd~ and synthetic spectra. The two upper panels show the synthetic spectra 
obtained when nickel is included in the spectrum only, thus in an inconsistent
manner. In the lower panels, nickel was included from the beginning in
the model atmosphere. It is possible to see the differences arising from
the two ways of treating the presence of nickel in the models. }
\end{figure}

\section{ABUNDANCE ANALYSIS}

\subsection{Observational material}

\subsubsection{FUV and UV Observations}

A wealth of FUV and UV observations of \bd28 has been gathered over the years. As we have mentioned earlier, \bd28 has been used as a calibration star by several space observatories. In order to find all the FUV and UV observations of \bd28, we carefully searched the Mikulski Archive for Space Telescopes (MAST)\footnote{http://archive.stsci.edu/index.html}. Spectroscopic observations of \bd28 have been obtained not only with space observatories such as {\it IUE}, {\it FUSE}, and {\it HST}, but also with instruments such as the {\it Orbiting Retrievable Far and Extreme Ultraviolet Spectrometers}  and the {\it Hopkins Ultraviolet Telescope} that flew on space shuttle missions. Moreover, the {\it HST} spectroscopic observations include observations obtained with the Goddard High-Resolution Spectrograph and STIS. After retrieving the \bd28 data from MAST and looking at the available data, we selected the {\it FUSE} and STIS observations in order to carry out our spectroscopic analysis. The choice of these spectroscopic data is based on the wide wavelength range covered by both instruments, and on their higher resolution and higher signal-to-noise ratio than the data obtained by other instruments. The next two sections describe in detail the {\it FUSE} and STIS data that we selected for our analysis.

\begin{deluxetable*}{lcccr}[th!]
\tablecolumns{5}
\tablewidth{0pt}
\tablecaption{STIS Datasets Retrieved From StarCAT
\label{tab_stis_obs}}

\tablehead{
\colhead{Dataset} &
\colhead{Grating} &
\colhead{$\lambda_{\rm{cen}}$ (\AA)} &
\colhead{Range (\AA)} &
\colhead{Exp. Time (s)} 
}
\startdata
E140M-1425\_020X020\_50710-53135 & E140M & 1425 & 1140.5--1730.0 & 20607 \\
E140H-1416\_020X020\_50812-53135 & E140H & 1416 & 1325.1--1505.8 & 15311 \\
E230M-1978\_020X020\_50710-53135 & E230M & 1978 & 1611.1--2150.1 & 9510  \\
E230H-2263\_020X020\_50812-53135 & E230H & 2263 & 2139.5--2394.0 & 22962 \\
E230H-2513\_020X020\_52461       & E230H & 2513 & 2386.5--2639.6 & 2674  \\
E230M-2707\_020X020\_50678-53135 & E230M & 2707 & 2629.6--3118.2 & 22547 \\
\enddata
\end{deluxetable*}

\subsubsection{{\it FUSE} Observations}

{\it FUSE} covers a wavelength range of 905~\AA\ to 1187~\AA\ with a resolution of about $R = \lambda/\Delta\lambda = 20$,000. For more information about the design of the instrument and the spectroscopic data products, see \citet{moos00}, \citet{sahnow00}, and \citet{dixon07}. During the lifetime of the {\it FUSE} mission, several observations of \bd28 have been carried out under three calibration programs. The first {\it FUSE} observation of \bd28 was obtained through the program M108 in order to establish the suitability of the star to the calibration program, given that its FUV flux was close to the bright limit of the {\it FUSE} detectors.  Subsequently, \bd28 was observed under the program M104 to test the use of the focal plane splits (FP splits) in order to mitigate the fixed-pattern noise, and therefore to increase the S/N ratio of the final spectrum. This technique consisted in taking a series of exposures at different focal plane assembly positions. In this way, the corresponding spectra were shifted along the dispersion direction and exposed on different portions of the detectors. The fixed-pattern noise was therefore reduced by co-aligning and co-adding all the spectra. The M104 observations consisted of exposures using the LWRS and MDRS apertures. Finally, two series of short observations taken through the LWRS, MDRS, HIRES apertures were carried out under the M103 program, which monitored the photometric stability of the {\it FUSE} instrument. 

We selected the M1080901, M1040101, M104105, M1031201, and M1031204 observations. All these observations were taken through the LWRS aperture and recorded in spectral image mode, or histogram (HIST) mode. We considered 66 exposures with an average exposure time of about 485 s, such that the total exposure time amounts to about 32,000 s. The eight {\it FUSE} segments, LiF1A, LiF1B, LiF2A, LiF2B, SiC1A, SiC1B, SiC2A, and SiC2B were cross-correlated and co-added. All LiF1B exposures show a depression in flux between 1130~\AA\ and 1170~\AA. This depression in flux is caused by a electron repeller wire grid that can cast shadows on the detectors. The LiF1B data in this wavelength range were not considered.  A final co-added {\it FUSE} spectrum was obtained by considering and merging the following spectral regions: SiC1B(905--990 \AA), LiF1A (990--1080 \AA), SiC2B (1080--1090 \AA), LiF2A (1090--1180 \AA), and LiF1B (1180--1187 \AA). The final spectrum has a signal-to-noise ratio S/N $\sim 68$ at 950~\AA,  S/N $\sim 130$ at 1050~\AA, and S/N $\sim 89$ at 1150~\AA. 

The {\it FUSE} spectrum displays many interstellar and stellar absorption lines. The most prominent stellar lines are the Lyman series of hydrogen starting from Ly$\beta$ up to Ly-8. There are followed by the \ion{He}{2} lines that involve transitions between energy levels $n_l = 2 \rightarrow n_u = 4$ up to 13. Transitions with upper even numbers produce, however, lines that are blended with the hydrogen Lyman lines.  The strongest metal lines are the \ion{O}{6} $\lambda\lambda$1031.91 and 1037.61 resonance lines that have equivalent widths of about 400~m\AA\ each. It is interesting to note that the \ion{O}{6} doublet does not display any P~Cygni profile characteristic, which suggests that no stellar wind expanding away from \bd28 is detected. In addition to the \ion{O}{6} lines, many metal lines of highly ionized species are observed. For instance, the following ions are observed in the {\it FUSE} spectrum: \ion{C}{4}, \ion{N}{4}, \ion{O}{4}, \ion{O}{5}, \ion{F}{4}, \ion{F}{6}, \ion{Si}{4}, \ion{P}{5}, \ion{S}{4}, \ion{S}{5}, \ion{S}{6}, \ion{Ar}{7}, \ion{Fe}{6}, \ion{Fe}{7}, \ion{Co}{6}, and \ion{Ni}{6}. The equivalent widths of the metal lines range from a few m\AA\ to $\sim$150~m\AA. Although a large number of metal lines have been identified, there are still about 150 absorption features with no identification. This state of affairs is also observed in FUV and UV spectra of many hot stars, and reflects the lack of atomic data for highly ionized species. Finally, the {\it FUSE} spectrum of \bd28 contains a large number of strong absorption lines from interstellar H$_2$, and also several absorption lines from species such as \ion{H}{1}, \ion{D}{1}, \ion{C}{1}, \ion{C}{2}, \ion{N}{1}, \ion{O}{1}, \ion{P}{2}, \ion{Ar}{1}, and \ion{Fe}{2}. \citet{son02} analyzed the line of sight toward \bd28 by using {\it FUSE} and STIS data, and measured interstellar \ion{D}{1}, \ion{H}{1}, \ion{N}{1}, \ion{O}{1}, \ion{Fe}{2} and H$_2$ column densities.

\subsubsection{STIS Observations}

Since the installation of STIS on board {\it HST} in 1997, \bd28 has been observed to monitor the sensitivity of each Multi-Anode Microchannel Array (MAMA) echelle grating mode. Although STIS suffered a major failure in August 2004, it returned to science operations after being repaired during the fourth {\it HST} servicing mission in May 2009. To this date about 206 observations of \bd28 have been carried out with the echelle gratings E140M and E140H FUV-MAMA, and the E230M and E230H NUV-MAMA. Specifically, \bd28 has been observed with the following echelle grating setups: E140M (central wavelength of spectrum $\lambda_{\rm{cen}} = 1425$~\AA; wavelength range 1140--1729~\AA), E140H (1416~\AA; 1316--1517~\AA), E230M (1978~\AA; 1607--2365~\AA), E230H (2263~\AA; 2128--2396~\AA), E230H (2514~\AA; 2385--2650~\AA), and E230M (2707~\AA; 2275--3118~\AA). Therefore the spectroscopic coverage ranges from 1140~\AA\ to 3118~\AA. The resolution $R=\lambda/\Delta\lambda$ for the E140M and E230M is $\sim45$,800, and it is $\sim114$,000 for the E140H and E230H.  All the observations were carried out by using the $0.2\arcsec \times 0.2\arcsec$ aperture. The exposure times for the individual observations are on average $\sim 360$~s and  $\sim 500$~s for the E140M and E230M observations, and they are $\sim1000$~s and $\sim1500$~s for the E140H and E230H observations. 

Instead of using all the STIS observations of \bd28 that are available at MAST, we opted to retrieve the observations from StarCAT\footnote{http://archive.stsci.edu/prepds/starcat/}, which is a STIS echelle spectral catalog of stars that \citet{ayres10} created based on observations of high-resolution spectra. StarCAT contains all the echelle high-resolution spectra of \bd28 that were collected from 1997 to the failure of STIS in 2004. Given that \bd28 has been observed frequently, \citet{ayres10} cross-correlated and co-added all the \bd28 spectra in order to achieve a high signal-to-noise ratio. We retrieved six datasets from \citet{ayres10}'s StarCAT. The properties of the datasets are summarized in Table~\ref{tab_stis_obs}. The table gives the name of the dataset, the grating that was used with its setting $\lambda_{\rm{cen}}$, the wavelength range, and the total exposure time in seconds that is the sum of all the exposures. The resulting signal-to-noise ratio for the observations taken with the E140M at $\lambda_{\rm{cen}} = 1425$ \AA\, is S/N $\sim 180$ at 1250~\AA\ and S/N  $\sim 145$ at 1500~\AA. The observations performed with the E230H at $\lambda_{\rm{cen}} = 2513$ \AA\  has the shortest exposure time and consequently has the lowest signal-to-noise ratio with S/N $\sim 30$ at 2500~\AA. The remaining observations have signal-to-noise ratios greater than 100. The resulting STIS spectra of \bd28 with their high signal-to-noise ratio and large spectral coverage are FUV and UV data of outstanding quality.

As in the case of the {\it FUSE} data, the STIS data show a large number of absorption lines. The greatest number of lines are observed between 1140~\AA\ and $\sim1475$~\AA. Beyond 1475~\AA, the intensity and the number of lines decrease dramatically. On the short wavelength side, absorption lines of high-ionization species such as \ion{Fe}{5}, \ion{Fe}{6}, \ion{Fe}{7}, \ion{Co}{6}, \ion{Ni}{5}, and \ion{Ni}{6} are the most numerous. The equivalent widths of these absorption lines vary from a few m\AA\ to $\sim 75$~m\AA. The strongest line observed in the STIS spectrum is Ly$\alpha$. In fact, most of the absorption at Ly$\alpha$ comes from the interstellar \ion{H}{1} along the line of sight of \bd28. \citet{son02} measured a \ion{H}{1} column density of $\log N({\rm{H\ I}}) = 19.842$ in the direction of \bd28, and showed that  the stellar \ion{H}{1} component is much fainter than the interstellar absorption. The second strongest observed line is the \ion{He}{2} $\lambda$1640 line ($n_l = 2 \rightarrow n_u = 3$) that has an equivalent width of about 1200 m\AA. The line shows broad wings that extend to about 4.5~\AA\ from the center of the line, and it shows a shallow core. Interestingly, the \ion{He}{2} line series $n_l = 3 \rightarrow n_u = 6$ up to 11 is observed at longer wavelengths. The wings of these \ion{He}{2} lines are also broad, but the lines show a much shallower core than the \ion{He}{2} $\lambda$1640 line. The \ion{N}{5} $\lambda\lambda$1238, 1242 lines and \ion{O}{5} $\lambda$1371 line are the strongest metal lines observed in the STIS spectrum. The lines have equivalent widths of about 770 m\AA, 482 m\AA, and 574 m\AA. The total equivalent widths of the \ion{Si}{4} $\lambda\lambda$1393, 1402 lines and \ion{C}{4} $\lambda\lambda$1548 and 1550 lines have equivalent widths of about 125~m\AA\ and 340~m\AA. 

Even though we have identified most stellar and interstellar absorption lines in the STIS spectrum of \bd28, there are still about 260 lines that do not have any identification. The equivalent widths of these lines range from a few m\AA\ to a few ten of m\AA. The total approximate number of lines with no identification in both {\it FUSE} and STIS spectra is around 410 lines. We extended our search of metal lines to elements beyond Zn. For instance, \citet{otoole06} and \cite{chayer06} observed strong absorption lines of heavy elements such as Ga, Ge, Zr, Sn, and Pb in {\it FUSE} and STIS spectra of sdB stars, while \citet{vennes05} and \citet{chayer05} observed Ge and Sn in a handful of hot DA white dwarfs, and Ge, As, Se, Br, Sn, Te, and I in two cool DO white dwarfs. \citet{werner12} added the discovery of Kr and Xe in the atmosphere of the DO white dwarf RE~0503$-$289 to the list of heavy elements detected in the atmospheres of compact stars. We looked for these heavy elements in both {\it FUSE} and STIS spectra by using the wavelengths of the high-ionization species. We also added Mo to our search. Unfortunately, no lines from heavy elements are observed. There is an absorption feature around 987.6~\AA\ that could correspond to the \ion{As}{5} $\lambda$987 line, but no absorption feature matches the second component of the \ion{As}{5} doublet at 1029.48~\AA\ adequately. As we have concluded in the previous section, the non-identification of many absorption features in the {\it FUSE} and STIS spectra of \bd28 illustrates the lack of atomic data of high-ionization species.

\subsection{Fitting Technique and Resulting Abundances}

\begin{figure*}[t!]
\begin{center}
\includegraphics[scale=.60,angle=270]{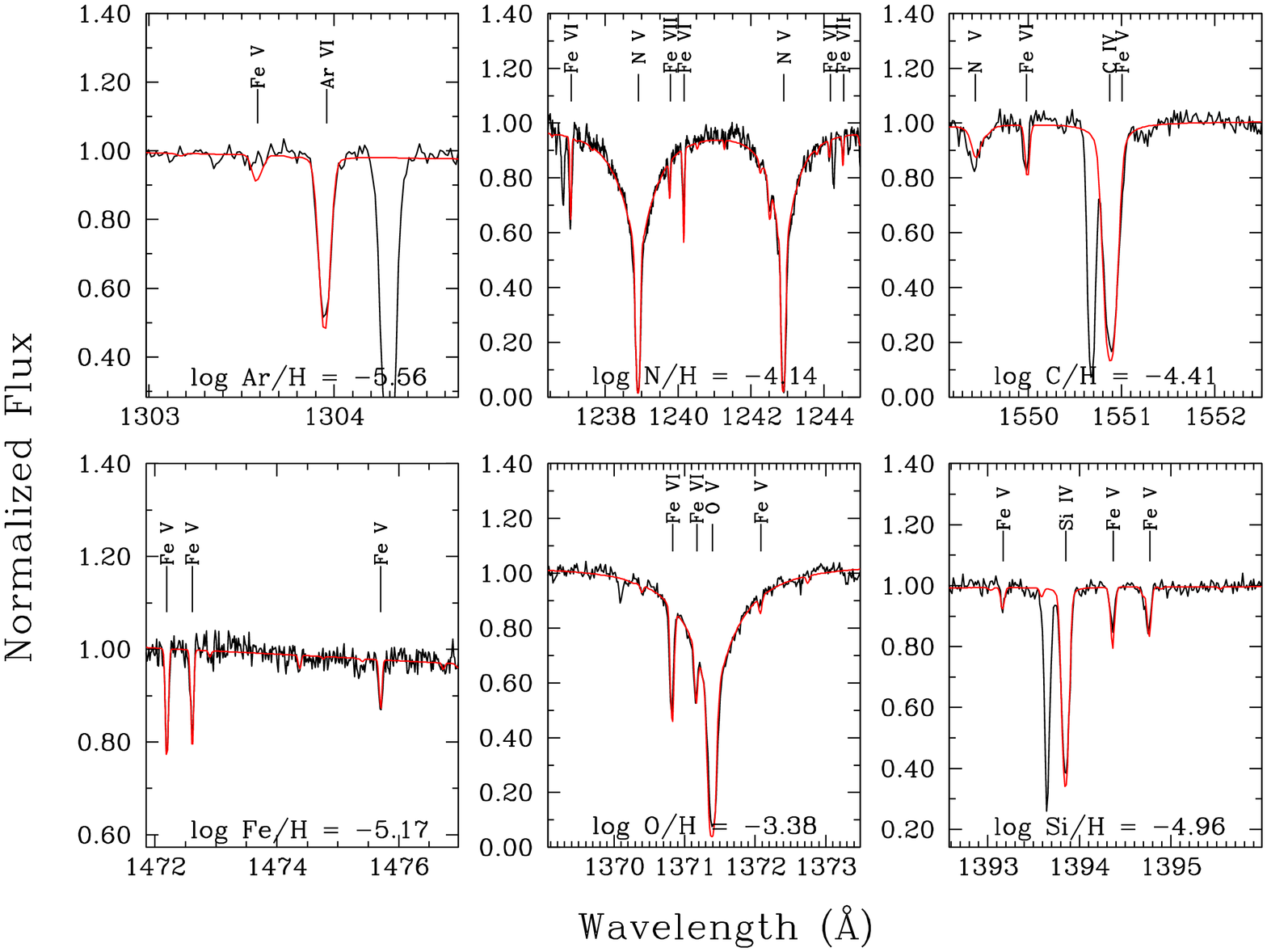}
\caption{Sample of our fitted spectral intervals from the
STIS spectrum. The red curve shows the result of the fitting procedure
for the element mentionned at the bottom of each panel, where the
resulting abundance for the interval is expressed as log
$N$(X)/$N$(H). All our models have \teff~= 82,000 K, log $g$ = 6.2, and
log \nhe~= $-$1.0. Interstellar features are visible in some spectral
chunks, such as a strong Si~\textsc{ii} line beside Ar~\textsc{vi}
$\lambda$1303 and shortward-shifted C~\textsc{iv} and Si~\textsc{iv}
lines. }
\end{center}
\end{figure*}

\begin{figure*}[th!]
\begin{center}
\includegraphics[scale=.60,angle=270]{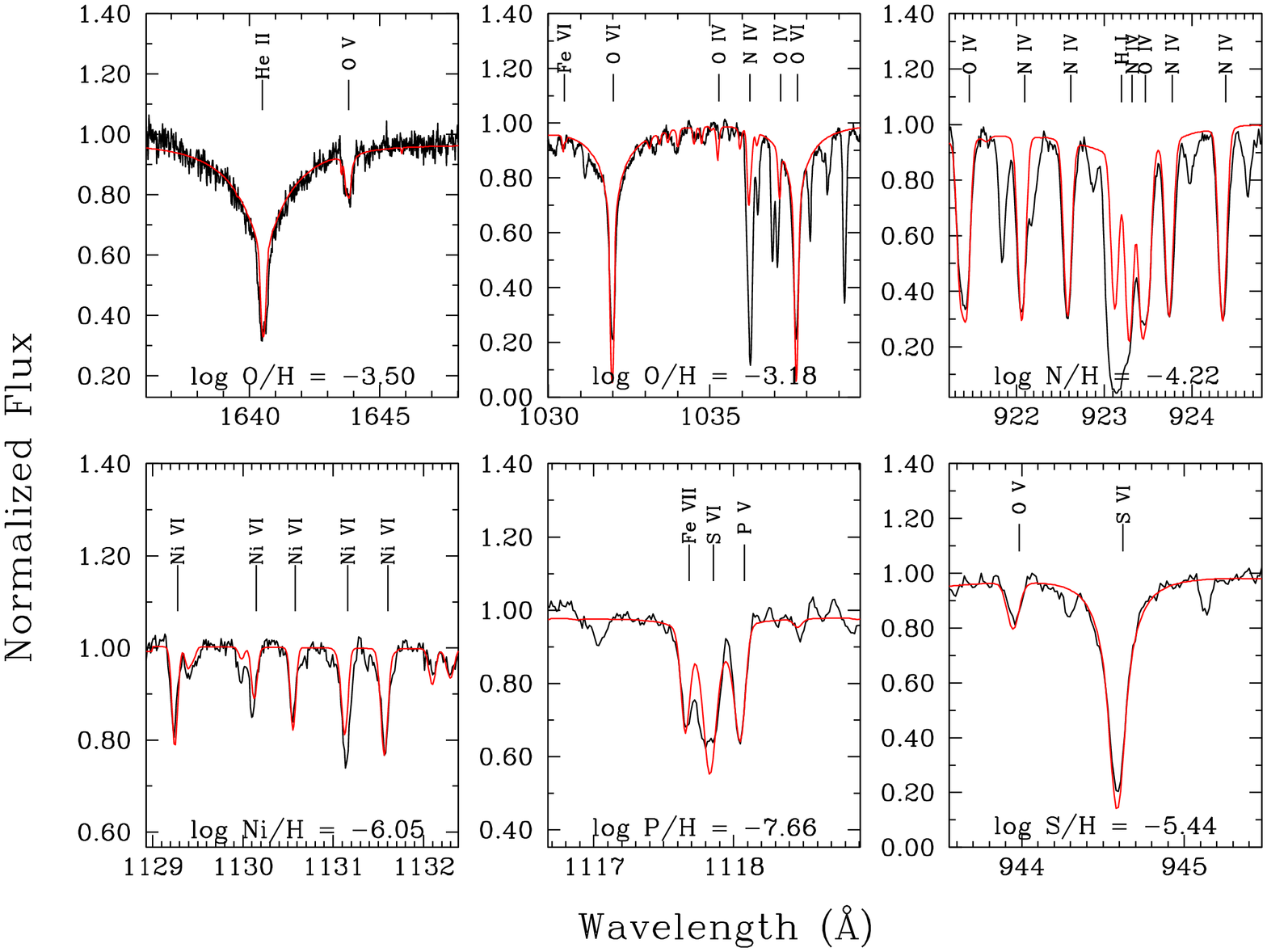}
\caption{Same as Figure 4 but for parts of the FUSE
spectrum, except for the first interval, showing He~\textsc{ii}
$\lambda$1640, which comes from the STIS spectrum. Note the
He~\textsc{ii} line which is particularly well reproduced with the
assumed fundamental parameters. There is a lot more interstellar lines
in the FUSE wavelength range. We notice in this figure, among others,
C~\textsc{ii} $\lambda$1036.3, O~\textsc{i} $\lambda\lambda$1039.19,
921.82 and C~\textsc{i} $\lambda$945.15. }
\end{center}
\end{figure*}

Our fitting technique consists in minimizing a $\chi^2$-type value
defined as the sum of the squared difference between the model and
observed fluxes over a given range of wavelength. Our free parameters
are the solid angle and, obviously, the abundance of the fitted
element. It is also possible to add a trend (linear or quadratic) if the
continuum in the range of interest needs one, which was sometimes useful
since the normalized UV spectra are not always as flat as they should. 

The method used to determine the abundance is a partial iterative
one. First of all, from the work of \citet{nap93}, the temperature and
gravity of our models are held fixed at \teff = 82,000 K and log $g$ =
6.2, and the helium abundance is fixed at the solar value.  Most of our
metal grids consist of six or seven different abundances centered around
the solar one and varying in steps of 0.5 dex. For example, in the case
of oxygen (whose solar abundance is log $N$(O)/$N$(H) = $-$3.3), its
grid covers a range from log $N$(O)/$N$(H) = $-$2.0 to $-$4.5 in steps
of 0.5 dex. Since we have a better idea of the iron and nickel
abundances (thanks primarily to the work of \citet{ram03}), their grid 
meshes are narrower ; between one tenth and one time solar for iron, and
centered around the solar value for nickel, but this time varying only
in steps of 0.2 dex. In the first step of the procedure, we used models
including C, N, O, and Fe in solar abundances initially and then varied
the abundances of these four elements, one at a time. We then used these
improved models to build the grids for silicon, sulfur, and phosphorus. 
We were then able to obtain a first estimate of the abundances for these
seven elements. The number of lines fitted for each element is given in
Table 2. We carefully chose lines that are reasonably well isolated in
regions of the spectrum not too crowded with blends or interstellar lines. 
Our second step was to redo our grids for each element, this time
including the previously found abundances for the other elements. At
this point, when redoing our fitting procedure, the values found for
individual lines sometimes changed a little, but the mean abundances
stayed roughly the same. Once we got the abundances of C, N, O, Si, P,
S, and Fe fixed, we made new grids in order to obtain, this time, a value
for nickel, fluorine, magnesium, and argon. The new grids for these four
elements only included, besides the element of interest, C, N, O and Fe,
which are the atomic species playing the major role in setting up the
atmospheric structure. We took a particular interest in argon because
the spectrum of \Bd~features two strong lines of this element, coming
from two ionization states : Ar~\textsc{vi} at 1303.86 \AA~and
Ar~\textsc{vii} at 1063.63 \AA. 

After these three steps, we were able to draw up a good portrait of the
chemical composition of \Bd, with quite satisfactory fits. Our final
results are summarized in Table 2.  The first column shows the different
elements as well as the ionic species present in our sample of fitted
lines. We then give the number of fitted intervals (like the six ones
presented in Figure 4) and the total number of lines (from the element
of interest) included in those intervals.  The third column presents the
mean abundance of the analyzed ranges while the fourth one gives the
standard deviation associated with the previous column. Finally, the
last one gives the total uncertainty of our abundances, which will be
discussed in the next subsection. 

We tentatively tried to formally fit some fluorine lines, but since they
were either too faint, or blended the results were not conclusive. Our
result is thus based on a sole isolated line, F~\textsc{iv}
$\lambda$1059.719, for which we visually estimated an abondance of log
$N$(F)/$N$(H) = $-$8.0 to be appropriate. The resulting comparison can
be found in the last panel of the online Figure. This result is 
compatible with the other lines we checked ($\lambda\lambda$ 1082.345,
1088.400 and 1139.523), although these lines are either blended or in a noisy
region of the spectrum for which their faintness does not help.  

For the sake of completeness, we also examined the chromium, manganese and cobalt lines 
visible in our spectra and estimated their abundances. This time we had to 
add these elements afterward in the synthetic spectrum (as explained in
section 2.2) where their populations were computed assuming LTE. The model
atmosphere used as input for the spectra computation included our main 
metallic elements (C, N, O, Si, P, S, Fe, and Ni). The abundances of the three
studied elements were in turn varied in the synthetic spectra and then fitted
the same way we did for the other elements. In order to avoid the ionization
problem discussed in section 2.2, we fitted only lines coming from the 
dominant ionization stage, which is~\textsc{vi} for these three iron-peak 
elements. As for the previously fitted elements, the uncertainties include
the standard deviation as well as effects from a change of temperature and
surface gravity in the input model atmosphere. 
Despite the fact that our resulting abundances presented in Table 3 are only
rough estimates, obtained with an approximate method, our values for Cr and Mn
are in good agreement with the ones found by \citet{ram03}, which were around
$-$5.88 for Cr and $-$6.62 for Mn.

Figures 4 and 5 show a sample of our fitted intervals taken from the
STIS (Figure 4) and FUSE (Figure 5, except the top left panel) spectra. The
abundance resulting from the fitting procedure of the featured region is
given as log $N$(X)/$N$(H). We show here the final step of our fitting
method, so the elements included, besides the fitted one, have 
abundances corresponding to the ones presented in Table 2. The totality of
our fitted spectral chunks are available as online material. 

Some of our intervals include important interstellar lines, such as
O~\textsc{i}, N~\textsc{i}, C~\textsc{i}, H~\textsc{i} and
H$_2$. Interstellar shortward-shifted C~\textsc{iv} and Si~\textsc{iv}
components are also seen for both doublets lines (one of which is shown
in Figure 4). These features are thought to originate from
a circumstellar cloud or shell near the star \citep{bru83}. Since
\Bd~has some similarities with central stars of old planetary nebulae in
its spectrum as well as in its fundamental parameters, this could be
some sort of old planetary nebula remnant.  When strong ISM lines are
present in our fitted intervals, we exclude them from our minimization
procedure by iteratively rejecting wavelength points of the observed
spectrum too far from the model's ones. 

We carefully inspected each fitted range to make sure the resulting model
was appropriate, the continuum was at a satisfactory level, and to notice any
discrepant lines.  Our checkup highlighted some points we found worth
mentionning.

\begin{deluxetable*}{lcccc}
\tablewidth{0pt}
\tablecaption{\Bd~ Chemical Composition --- Results of our abundance analysis}
\tablehead{
\colhead{Element} &
\colhead{Nb of intervals} &
\colhead{Mean abundance} &
\colhead{Standard deviation} &
\colhead{Total uncertainty}\\
\colhead{(ions)} &
\colhead{(Nb of lines)} &
\colhead{log $N$(X)/$N$(H)} &
\colhead{(dex)} &
\colhead{(dex)} 
}
\startdata
C (\textsc{iv})  & 3 (3)  & $-$4.48  & 0.16  & 0.46  \\ 
N (\textsc{iv}, \textsc{v})  & 4 (8)  & $-$4.23  & 0.19  & 0.78  \\
O (\textsc{iv}, \textsc{v}, \textsc{vi})  & 10 (15)  & $-$3.48  & 0.15  & 0.46  \\
F (\textsc{iv}, \textsc{v}, \textsc{vi})\tablenotemark{a} & 1 (1)  & $-$8.00  & --  & 0.50 \\
Mg (\textsc{iv}) & 7 (7) & $-$4.57  & 0.12  & 0.45 \\ 
Si (\textsc{iv})  & 3 (3)  & $-$4.95  & 0.06  & 0.30  \\
P (\textsc{v})  & 2 (2)  & $-$7.45  & 0.29  & 0.52  \\
S (\textsc{v}, \textsc{vi})  & 8 (8)  & $-$5.53  & 0.18  & 0.47  \\
Ar (\textsc{vi}, \textsc{vii})  & 2 (2)  & $-$5.53  & 0.04  & 0.43  \\
Fe (\textsc{v}, \textsc{vi}, \textsc{vii})  & 11 (33)  & $-$5.08  & 0.12  & 0.32  \\
Ni (\textsc{v}, \textsc{vi})  & 7 (17)  & $-$6.04  & 0.21  & 0.48  \\
\enddata
\tablenotetext{a}{Fluorine was not formally fitted, we visually examined some lines in order to estimate an abundance. See the text for more details.}
\end{deluxetable*}

\begin{deluxetable*}{lcccc}
\tablewidth{0pt}
\tablecaption{Chromium, manganese and cobalt estimated abundances (LTE)}
\tablehead{
\colhead{Element} &
\colhead{Nb of intervals} &
\colhead{Mean abundance} &
\colhead{Standard deviation} &
\colhead{Total uncertainty}\\
\colhead{(ions)} &
\colhead{(Nb of lines)} &
\colhead{log $N$(X)/$N$(H)} &
\colhead{(dex)} &
\colhead{(dex)} 
}
\startdata
Cr (\textsc{vi})  & 4 (4)  & $-$6.17  & 0.31  & 0.41 \\ 
Mn (\textsc{vi})  & 5 (6)  & $-$6.71  & 0.21  & 0.32  \\
Co (\textsc{vi})  & 5 (11) & $-$6.71  & 0.12  & 0.25  \\
\enddata

\end{deluxetable*}

First of all, if we consider the most prominent features (i.e., resonance
and strong lines), they are all well reproduced. The only exception is
the O~\textsc{vi} doublet ($\lambda\lambda$1032, 1038), for which the
cores happen to be too opaque in our models. This phenomenon can also be
seen, at a smaller amplitude, in the cores of the resonance doublet of
S~\textsc{vi} ($\lambda\lambda$933, 944), in S~\textsc{vi}
$\lambda$1117.76 and in O~\textsc{v} $\lambda$1371.3.  We have to
mention here that the analysis of \citet{rauch07} has been a useful
reference to us on this particular account.  They analyzed STIS and FUSE
spectra of a star (LS~V~$+$46$\arcdeg$21) whose fundamental parameters
(\teff~= 95,000 K and log $g$ = 6.9) give it a spectrum quite similar to
that of \Bd, in the sense that both star show lots of common lines. It
allows us to compare some of our fits with theirs. This way, we noticed
that they reported the same effect of too strong cores in resonance
lines of their star, letting us know the issue is not only about our own
model atmospheres.   

Secondly, when choosing our sample of lines to be fitted, a few lines
gave inconsistent results or could not be matched in any way
whatsoever. For instance, our attempt to fit the O~\textsc{iv} structure
around 1081 \AA~was not conclusive enough to be included in our final
sample because the core of the central component at 1080.97 \AA~was too
strong in our models. The N~\textsc{iv} triplet between 1225 and 1226
\AA~required a much higher abundance (log $N$(N)/$N$(H) = $-$3.2) than
our other lines of nitrogen, so we did not include it our sample. The
same thing happened for O~\textsc{v} $\lambda$968.9, but this time we
suspect this line to be blended with interstellar H$_2$ lines at 968.997
and/or 969.07 \AA. This possibility is supported by the few H$_2$ lines
present in the vicinity of the oxygen line. 

Finally, some odd lines of oxygen appear in our models, whereas there is
no sign of them in the observed spectrum. We noticed
O~\textsc{iii}~$\lambda$1153.775 and two O~\textsc{iv} lines at 1076.06
and 1294.065 \AA.  A too strong oscillator strength value might be the
cause of their presence. We should also mention four other lines, among
which three of them appear also too strong in our final model, namely
Fe~\textsc{v} $\lambda$1393.072 and Fe~\textsc{vii}
$\lambda\lambda$1154.990, 1180.827. As for the last one, Ni~\textsc{vi}
at 1204.078 \AA~is too faint in our models and was excluded from our
fitted nickel sample. However, these lines also appear too strong (or
too faint for the nickel one) in the final model of \citet{rauch07}, so
there is a strong chance that the oscillator strengths of the lines are
involved here. 

\subsection{Evaluation of the Abundance Uncertainties}

Hot stars sometimes ``become'' hotter with time, in the sense that their
estimated effective temperature tend to be revised upwards. For example,
the DAO white dwarf LS~V~$+$46$\arcdeg$21, has seen its estimated effective
temperature go from 83,000 K in \citet{nap99} to 95,000 K in the
thorough study of \citet{rauch07}. In a most extreme case, the white
dwarf KPD 0005$+$5106 was shown to be the hottest known DO in
\citet{wer94} with an estimated value of \teff~= 120,000 K, but a
subsequent analysis of better spectroscopic data found lines of highly
ionised metals (Ne~\textsc{viii} and Ca~\textsc{x}), thus needing the
star to have an effective temperature of at least 180,000 K
\citep{wass10}. With these rather extreme cases in mind, we wanted to
have an idea of how our abundances would change if the effective
temperature or the surface gravity of \Bd~end up being different 
from our assumed values. 

Therefore, we redid a part of our abundance analysis, using in a first
step models with an effective temperature of 92,000 K, and then models
at log $g$ = 6.6.  To do this for all the atomic species studied in the
previous section would require computing twice as many grids as we did
for determining the abundances themselves. We therefore decided to do
this exercise for three elements only, namely nitrogen, silicon, and
iron. The resulting abundances are presented in Table 4 which gives, for
each model considered, the abundance and standard deviation obtained for the
indicated element. The first column specifies the model used in our
fitting procedure by indicating the parameter that has been changed with
respect to the ones used in the previous section. The first quoted model
is our reference model at \teff~= 82,000 K and log $g$ = 6.2.  One thing to
note here is that a change in the effective temperature of the models of
the magnitude considered here induces a larger change in the determined
abundance than a change in log $g$.  Another interesting thing to look
at is the value of $\sigma$, the standard deviation of the different
spectral chunks we fitted. In the case of iron and nitrogen, this value
is larger for \teff~= 92,000 K and 72,000 K, meaning the abundances
obtained from each interval agree less with each other than in the
reference case. The results for silicon are somewhat different because
there is only three fitted lines, produced by a sole ion
(Si~\textsc{iv}), whereas nitrogen and iron have lines produced by two
and three ions.  Silicon is thus less expected to show 
larger discrepancies between its fitted lines at different atmospheric
parameters. The behavior of the standard deviation (for nitrogen and
iron) with the temperature can be taken at a reassuring sign pointing
towards 82,000 K to be a good value for the star's effective
temperature.  About the two additional log $g$ values, their standard
deviation for iron is a bit larger than the one found in the previous
section but as a whole, they are nevertheless quite similar to the ones
obtained with models at log $g$ of 6.2.

We used the abundance differences between different models in the
calculations of the uncertainties reported in Table 2. We wanted our
chemical composition, within the given uncertainties, to be able to
stand a potential change in the fundamental parameters of \Bd. By using
the abundances found at \teff~=92,000 K and log $g$ = 6.6, our
determined values should remain appropriate for changes of $\pm$ 10,000
K and $\pm$ 0.4 dex, assuming the errors should be symmetrical in
\teff~and log $g$. Although we also have abundances for 72,000 K and log
$g$ of 5.8 we prefer not to include them in our uncertainty calculations
because they are less likely to be realistic values for \Bd. This will
be discussed in the next section.  That being said, our final
uncertainties are the sum of three components :  
\begin{equation}
\sigma =\sqrt{(\sigma _{\rm{fit}})^2+(\sigma _{\rm{Teff}})^2+(\sigma
  _{\rm{log}g})^2}, 
\end{equation}
 
\noindent where $\sigma _{\rm{fit}}$ is the standard deviation of the fitted
chunks, $\sigma _{\rm{Teff}}$ is the difference between the abundances
at \teff~=92,000 K and 82,000 K while $\sigma _{\rm{log}g}$ is the
difference between the results at log $g$ of 6.6 and 6.2.  For the
elements besides nitrogen, silicon and iron, $\sigma _{\rm{Teff}}$ and
$\sigma _{\rm{log}g}$ were taken as the mean values of the three
determined ones.

\begin{deluxetable*}{lcccccc}[b]
\tablewidth{0pt}
\tablecaption{ Determined abundances of nitrogen, silicon and iron}
\tablehead{
\colhead{Models} &
\multicolumn{2}{c}{Nitrogen} &
\multicolumn{2}{c}{Silicon} &
\multicolumn{2}{c}{Iron} \\
\colhead{} &
\colhead{Abundance} &
\colhead{$\sigma$} &
\colhead{Abundance} &
\colhead{$\sigma$} &
\colhead{Abundance} &
\colhead{$\sigma$} 
}
\startdata
\teff~82 kK  & $-$4.23  & 0.19  & $-$4.95  & 0.06  & $-$5.07  & 0.12 \\
\teff~92 kK  & $-$3.47  & 0.32  & $-$4.66  & 0.05  & $-$4.83  & 0.58 \\
\teff~72 kK  & $-$4.64  & 0.33  & $-$5.32  & 0.13  & $-$5.03\tablenotemark{a}  & 0.35\tablenotemark{a} \\
log $g$ 6.6  & $-$4.24  & 0.16  & $-$4.95  & 0.02  & $-$4.89  & 0.20 \\
log $g$ 5.8  & $-$4.15  & 0.18  & $-$4.94  & 0.13  & $-$5.23  & 0.14 \\
\enddata
\tablenotetext{a}{The quality of the iron fits with these models is rather poor.}
\end{deluxetable*}

\section{CONSTRAINING THE ATMOSPHERIC PARAMETERS}

\subsection{With the Metal Lines}

As mentionned in the Introduction, finding out the effective temperatures
and surface gravities of hot stars is not straightforward and
standard methods used on cooler stars do not work very well. A
more reliable approach in this case is to look at the metal lines
visible in the UV spectrum of the star. A change of temperature will
modify the ionization equilibrium of the atomic species present in the
photosphere, as shown previously in Figure 2, and this will result in
changes of the spectral lines strength. The exercise done in the last
subsection, when fitting iron and nitrogen with models having different
parameters, favors the models at \teff~= 82,000 K. The standard
deviations of the fitted intervals are smaller with this temperature and
moreover, the ionization equilibrium of iron lines cannot be correctly
reproduced neither with the hotter or cooler models. This is what is
shown in Figure 6, where the two left panels feature an interval of the
STIS spectrum including three ionization stages of iron and three model
spectra having different temperatures. Fe~\textsc{vi} lines do not
change much with the effective temperature, but Fe~\textsc{v} and
\textsc{vii} lines are more sensitive and become quite stronger with,
respectively, a lower and higher temperature. The changes are easily
seen, even with a difference of 5,000 K between models.  Thus, when
trying to do a fit of the iron lines, at hotter or cooler temperatures,
it is impossible to simultaneously match the lines coming from the three
ionization degrees. Indeed, the four fits shown in Figure 6 (middle and
right panels) are rather poor. Therefore, when looking at iron lines as
a temperature indicator, it appears clearly that the temperature of the
star must be quite close to 82,000 K. 

\begin{figure*}[t!]
\begin{center}
\includegraphics[scale=.60,angle=270]{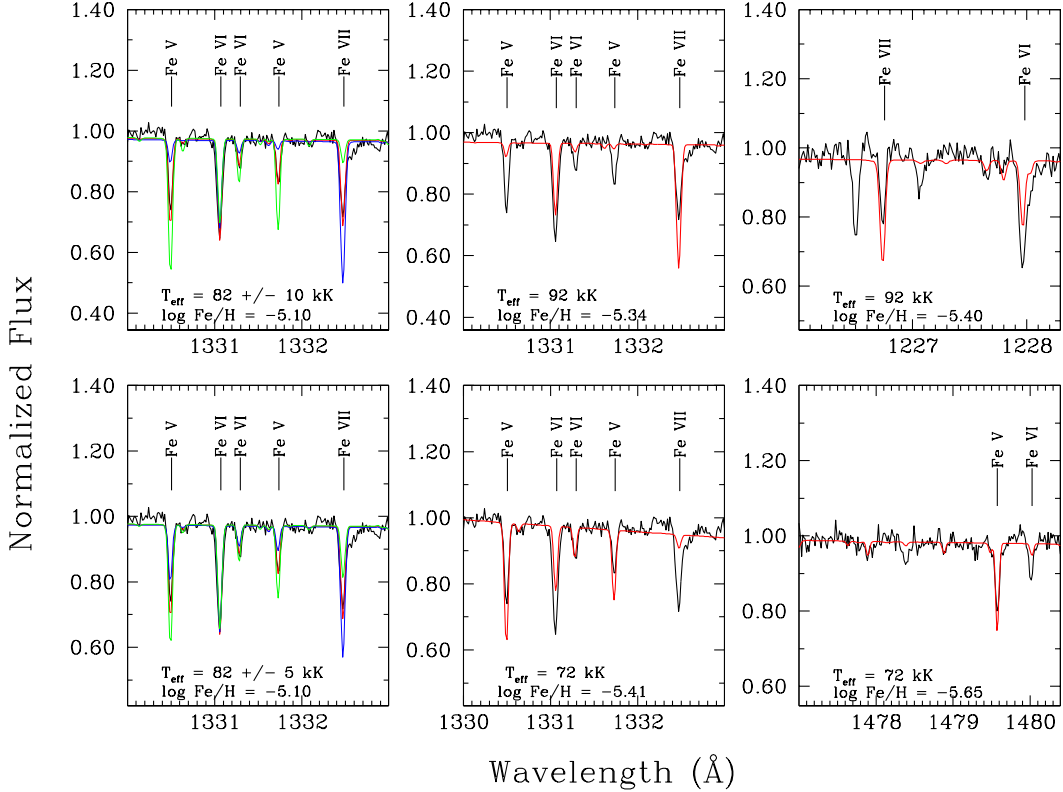}
\caption{Comparison of iron lines with models having
different temperatures. The temperature and the abundance of iron are
indicated on each panel. The two left panels present a comparison of
three model spectra with the spectrum of \Bd~over a range featuring
lines from three ionization stages of iron. The standard model at
\teff~=~82,000 K and log $g$ = 6.2 is represented by the red line, the
hotter ones (92,000 K in the top panel and 87,000 K in the bottom one)
are in blue, while the cooler models (72,000 K in the top panel and
77,000 K in the bottom one) are in green. It can be seen that with a
fixed abundance, a change in the temperature, even of only 5,000 K, leads
to a poorer agreement between the observed lines and the modeled ones. 
The two middle panels show the result of a fitting procedure, over the
same wavelength range, with hotter and cooler models, while the right
panels also show a fit of iron lines, but over two other spectral ranges.  }
\end{center}
\end{figure*}

\begin{figure*}[t!]
\begin{center}
\includegraphics[scale=.60,angle=270]{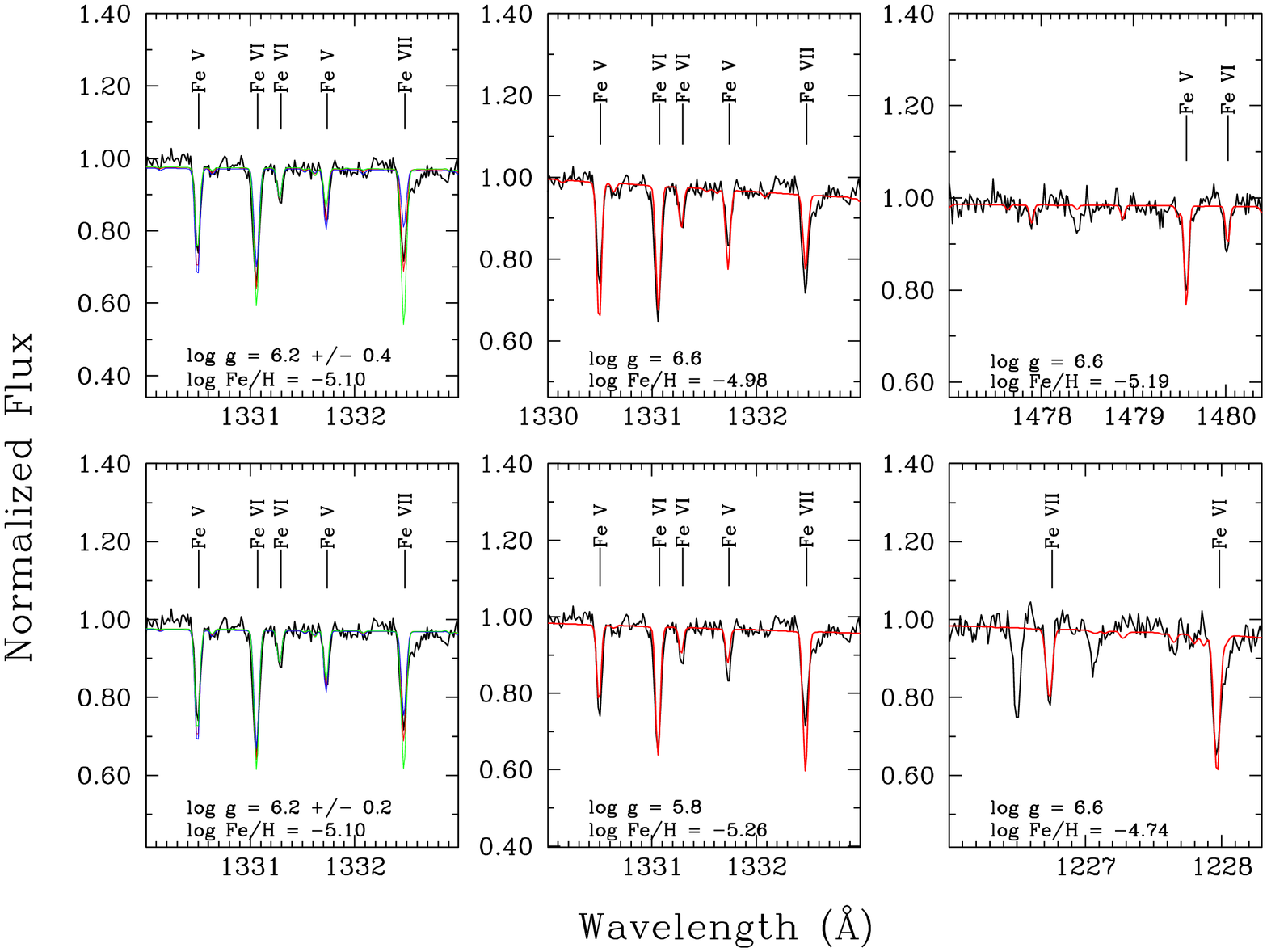}
\caption{ Comparisons of iron lines with models at different
log $g$. This figure is very similar to Figure 6, but for different
values of log $g$ while the temperature is kept at 82,000 K. The two
left panels show the observed spectrum and three models with different
surface gravity, having a fixed iron abundance. Models in the upper
panel have gravities that change by 0.4 dex with log $g$ = 6.2 as a
middle value, while the changes are of 0.2 dex in the lower panel. The
spectrum with the highest value is in blue, the middle one in red, and
the lowest one in green. The two middle panels show the results of the fits
with the two extreme values of log $g$, while the right panels show
different ranges, fitted with models at log $g$ of 6.6.
}
\end{center}
\end{figure*}

Figure 7 shows the same kind of plots than Figure 6, except that this
time we changed the value of the surface gravity between log $g$ of 6.6
and 5.8 while the temperature was fixed at \teff~= 82,000 K. Like the
temperature, the gravity also affects the iron lines, but to a lesser
extend, at least in the ranges investigated. With a higher log $g$ it is
still possible to represent correctly the lines shown in the two right
panels, but the fits lead to abundances that are different by 0.45 dex,
which is larger than the difference of 0.12 dex obtained with our models
at log $g$ = 6.2 (see the online Figure, panel g). When fitting our
principal range of interest, 1330 $-$ 1333 \AA, with models having
different log $g$, we do not obtain a match as good as with our
reference grid. Indeed, when looking at the standard deviations for the
iron fits found in Table 4 for the various log $g$, the agreement is
better with the reference model, but $\sigma$ is not much higher in the
two other cases. In the case of nitrogen, the standard deviations of the
various log $g$ do not obviously support a specific gravity. That being
said, when considering iron lines, the surface gravity we assumed for
our analyses (log $g$ = 6.2) seems to be right. 

\subsection{With the Parallax Distance}

Another method that can be used to place constraints on the parameters
of \Bd~is to compare its spectroscopic distance with the one determined
by parallax measurements, which is between 81 and 106 pc according to
the latest reduction of the $Hipparcos$ catalogue \citep{hip07}. The idea
here is to compute the absolute magnitude of a model atmosphere and combine
it with the apparent magnitude of the star (recently mesured by
\citet{land07}) to get the distance. In order to do that, some other
quantities are required.  

First of all, at these distances, reddening must be considered, and
we thus computed theoretical, unreddened $(B-V)_o$ color indices for
several relevant model spectra. We used the flux calibration of
\citet{hol06} to find the absolute $B$ and $V$ magnitudes. By considering
six models (with \teff = 82,000 K and 87,000 K, and log $g$ = 6.2, 6.4,
and 6.6), we derived an average representative color index $(B-V)_o =
-0.3831 \pm 0.0018$ showing a very small dispersion (which is, of course,
not surprising for a star as hot as \Bd). This is to be compared with
the accurate observed value of $(B-V) = -0.3410 \pm 0.0018$ obtained
byn\citet{land07}, leading immediately to a rather precise reddening
index of $E(B-V) = 0.042 \pm 0.003$ for \Bd. Combined with the
extinction law proposed by \citet{sea79}, this leads to an absorption
coefficient in the $V$ band of $A_V = 3.20 E(B-V) = 0.135 \pm 0.008$. We
note that this value is quite compatible with the overall absorption
coefficient of $A_V = 0.3$ along the line-of-sight in the direction of
\Bd~as obtained from the Infrared Science 
Archive\footnote{http://irsa.ipac.caltech.edu/applications/DUST/} using 
data from \citet{dust98}. 

Another parameter needed in the computation of the spectroscopic
distance is the mass of the star, which allows to  find its radius given
the surface gravity. Unfortunately, that parameter is largely unknown,
except to say that \Bd~must be either a post-AGB, a post-EHB, or maybe
a post-RGB (though this is less likely because of the relatively short
timescale of this evolutionary path). Constraints on the mass can then
be derived from model calculations of these late evolutionary
phases. For instance, according to the evolutionary tracks of
\citet{scho83} (post-AGB), \citet{dor93} (post-EHB), and \citet{drie98}
(post-RGB) and the approximate position of \Bd~in the log $g$ $-$
\teff~plane, it would appear that its mass should be around 0.5
\msun~and could hardly be higher than 0.6 \msun~or lower than 0.4
\msun. Plots of post-AGB and post-EHB tracks can be found in Figure 1 of
\citet{haas96}, while post-RGB tracks are shown in Figure 10 of
\citet{stro07}. However, this is probably not the full story because we
know of post-EHB stars with masses less than 0.4 \msun, one of which not
known to be part of a close binary system (\cite{heb05};
\cite{ran07}; \cite{for10}; \cite{fon12}). We cannot therefore exclude a
mass for \Bd~less than 0.4 \msun~ and, as an extreme limit, we will also
consider a value as low as 0.3 \msun.

Finally, the surface gravity and the effective temperature of a given
model atmosphere influence direcly the absolute magnitude determined in
a given bandpass and, thus, the inferred distance. Since the
effective temperature of \Bd~appears to be well constrained by the
pattern of iron lines, this parameter was initially fixed at \teff~=
82,000 K and we computed the spectroscopic distance of the star for different
combinations of mass and surface gravity. This was done by comparing the
computed absolute visual magnitude $M_V$ with the well-measured reddened
apparent magnitude of $V = 10.509 \pm 0.0027$ provided by \citet{land07}

Our results are presented in Table 5, where, for each combination of mass
and gravity, we computed two distances, with and without the
reddening. The comparison of our computed spectroscopic distances and
the measured one favors a surface gravity higher than 6.2 and/or a low
mass for \Bd. If we were to insist that the mass of \Bd~ is a
representative post-EHB star value, 0.5 \msun~say, then our optimal
spectroscopic model -- characterized by \teff~= 82,000 K and log $g$ =
6.2 -- would lead to a distance of 157 pc, in apparent conflict with the
parallax measurement of 81$-$106 pc. This is very reminescent of the
situation encountered by \citet{rauch07} in the case of the hot DAO
white dwarf LS~V~$+$46$\arcdeg$21 where the authors estimated the
unknown mass by interpolating in a given set of evolutionary
tracks. With a fixed value of 0.55 \msun, they found a
discrepant spectrocopic distance of 224$_{-58}^{+46}$ pc compared to a
ground-based parallax measurement giving 129$_{-5}^{+6}$ pc. 

\begin{deluxetable}{ccccc}
\tablewidth{0pt}
\tablecaption{Spectroscopic distances (pc) obtained with models having
  \teff~= 82,000 K in $V$-band with A$_{\rm{v}}$ = 0 and 0.135} 
\tablehead{
\colhead{Mass / log $g$} &
\colhead{0.3 \msun} &
\colhead{0.4 \msun} &
\colhead{0.5 \msun} &
\colhead{0.6 \msun} 
}
\startdata
log $g$ = 6.2  & 129 / 122 & 150 / 140  & 167 / 157  & 183 / 172 \\
log $g$ = 6.4  & 103 / 97  & 119 / 111  & 133 / 125  & 145 / 136  \\
log $g$ = 6.6  & 81 / 77   & 94 / 88    & 105 / 99  & 115 / 108  \\
\enddata
\end{deluxetable}

If we take the parallax measurement of \Bd~at face value, then the
surface gravity has to be pushed above log $g$ $\sim$ 6.5 for our
spectroscopic distance to become compatible with that measurement, again
assuming that the mass of the star is 0.5 \msun. However, such large
values of the surface gravity are now in conflict with the iron line
profiles depicted in Figure 7. It may thus be preferable to think in
terms of a low mass for \Bd~for the time being. This may also be an
option in the case of LS~V~$+$46$\arcdeg$21 \citep{rauch07}.

Finally, we also checked what would be the effect of a change in the
temperature of our model spectra and we thus computed a table similar to
Table 5, except that all models were characterized by \teff~= 87,000 K
instead of 82,000 K. We found that the spectroscopic distance increases
by only 2 to 6 pc, depending on the parameters mass and gravity, compared
to the entries of Table 5. Hence, within an uncertainty of 5000 K (which
seems to be a reasonable range according to the result of the previous
section), the conclusions of this subsection are practically not
dependent on the effective temperature. 

\section{DISCUSSION}

In spite of the huge progress made in the model atmosphere modeling
field in the last two decades, the analysis of hot stars still remains a
challenge. In this paper, we have presented the first part of our
analysis of \Bd, which consists in the study of its UV spectrum.  Our
work made use of high-quality spectra from the $HST$ and $FUSE$
satellites combined with state-of-the-art NLTE line-blanketed model
atmospheres and synthetic spectra computed with TLUSTY and SYNSPEC. To
our knowledge, the $FUSE$ data available on that star have not been
exploited previously in the context of an atmospheric abundance analysis. 
The abundances of eleven elements have been determined, namely those of
C, N, O, F, Mg, Si, P, S, Ar, Fe, and Ni. Our abundance analysis was made in a
self-consistent way, meaning that the element analyzed contributed to
the thermodynamical structure of the model and its populations were
explicitly calculated in NLTE during the computation of the model
atmosphere. We also made sure that the models used for fitting elements
always included, besides the fitted element, at least the atomic species
that most contribute to the thermodynamical structure, that is to say
helium, carbon, nitrogen, oxygen, and iron. For a star as hot as \Bd, we
stress that it is crucial to compute NLTE populations of the studied
elements in order to get realistic abundances. Even with the thermodynamical
structure of a NLTE line-blanketed model atmosphere, the LTE ionization
equilibrium is likely to be wrong, thus preventing a simultaneous fit 
of absorption lines originating from different ionization stages of a given
atomic species. We illustrated this effect in Figure 3, with an example of 
nickel lines computed in both the NLTE self-consistent way
and with LTE populations. In addition to the elements mentionned above,
the UV spectrum of \Bd~shows lines of chromium, manganese, and cobalt 
(and probably those of other species as well). We tentatively tried to 
fit these three elements by using the LTE approximation, because of 
the lack of proper model atoms that could be used in our NLTE models.  
We analyzed only lines originating from
the dominant ionization stage (\textsc{vi}) and we got abundances 
surprisingly similar to the ones \citet{ram03} derived (for Cr and Mn) using 
the ``generic ion'' approach in NLTE model atmosphere.

\begin{figure}[b!]
\begin{center}
\epsscale{1.3}
\plotone{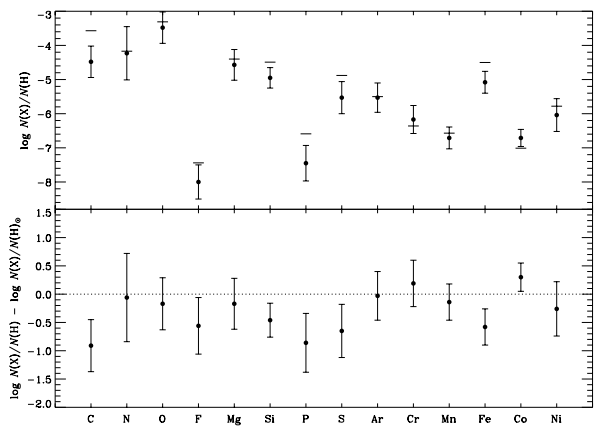}
\caption{Summary of the determined chemical composition of \Bd.
The top panel shows the absolute abundances relative to hydrogen (log $N$(X)/$N$(H))
and the horizontal lines indicate the solar one for each element \citep{asp09}. 
The bottom panel shows this time the abundances relative to the solar value.}
\end{center}
\end{figure}

Our resulting chemical composition is summed up in Figure 8 and
the entire sample of our NLTE fitted lines is shown in the online Figure. 
We found the overall quality of our fits very satisfying. When comparing
the abundances of \Bd~with those of the Sun, it appears that none of the
elements studied with self-consistent NLTE models has an abundance higher 
than solar (the only exceptions being the LTE estimated abundances of chromium
and cobalt).
Instead, our derived abundances are all between one and $1/10$ of the solar ones
 \citep{asp09}.
The most depleted species is carbon with log $N$(C)/$N$(C$_{\odot}$) =
$-0.91 \pm 0.46$, followed by phophorus with log $N$(P)/$N$(P$_{\odot}$) =
$-0.86 \pm 0.52$. It turns out that these two elements, as mentionned in
subsection 2.2, are present in the photosphere mainly as C~\textsc{v}
and P~\textsc{vi}, which are in a noble gas configuration. This kind of
ion usually has its resonance lines in the extreme UV (EUV) or even in
the X-ray domain because of the large energy gap between their ground
and first excited levels. Therefore, they are less sensitive to
radiative levitation, since the EUV and X-ray fluxes of stars are
usually lower than the flux at larger wavelengths. This might explain
why carbon and phosphorus are the most depleted elements, since their
resonance lines are respectively around 40~\AA~and 90~\AA. Even if
\Bd~is a hot star, its flux at these extreme wavelengths remains quite
low. For its part, even though our abundance of fluorine is at best an 
estimation based on a sole line, its value around 
log $N$(F)/$N$(F$_{\odot}$) = $-0.56 \pm 0.50$ still suggests that this
element is not enriched in \Bd~. This is in line with the findings of 
\citet{wer05} who found an important enrichment of fluorine in a number
of PG1159 stars, while its abundance remains around or slighly lower
than solar in their sample of H-rich central stars of planetary 
nebulae, the latter being a family of stars having similar evolutionary paths
than \Bd.  As for iron, our results indicate that log
$N$(Fe)/$N$(Fe$_{\odot}$) = $-0.58 \pm 0.32$, while they indicate 
log $N$(Ni)/$N$(Ni$_{\odot}$) = $-0.26 \pm 0.48$ for nickel. Both
results are compatible with the findings of \citet{ram03} which suggest
a depleted value of iron by about an order of magnitude (formally, from
their Table 1, log $N$(Fe)/$N$(Fe$_{\odot}$) = $-0.82 \pm 0.15$), and a
solar abundance for nickel\footnote{Our derived abundances log
  $N$(N)/$N$(N$_{\odot}$) = $-0.06 \pm 0.78$ and log
  $N$(O)/$N$(O$_{\odot}$) = $-0.17 \pm 0.46$ are equally compatible
  with the work of \citet{haas96} who deduced about solar abundances for
  those two elements.}. We thus find a ratio $N$(Fe)/$N$(Ni)
$\simeq$ 9.1, which is about half of the solar value. An even more depleted
ratio was first suggested by \citet{haas96} in \Bd~and is particularly
interesting because it connects well with the work of \citet{wer94ni} 
who found a similar trend in four hot DA white dwarfs. This trend may
possibly be explained through the combined effects of radiative
levitation and residual stellar winds (see, e.g., the work of
\citet{cha94}). Both elements have similar atomic masses and structures,
but the lower ratios obtained in hot stars suggest that nickel is more
sensitive to radiative levitation and/or winds than iron, thus 
allowing a larger proportion of nickel to stay in the photosphere. It
must be kept in mind that with an effective temperature of 82,000 K, the
abundance pattern of \Bd~is strongly affected by radiative
levitation. We made sure that the uncertainties on our determined
abundances were computed in a way that includes the effects of a change
in the temperature and the gravity of our model atmospheres. Therefore,
our abundances should remain fairly reliable even if the effective
temperature or the surface gravity are revised upward or downward by
some 10,000 K or 0.4 dex, respectively \footnote{After our complete
analysis, it appears that a $\pm$10,000 K margin of error on the effective
temperature is twice our adopted value. Thus, the uncertainties on our
abundances are a bit generous.}.

Even if we took the fundamental parameters of \Bd~from the literature
\citep{nap93}, we nevertheless checked the validity of the assumed
effective temperature and surface gravity. To carry out that task, we
first redid our fits for nitrogen, silicon, and iron (used 
as proxy elements) by constructing additional model atmospheres having
different parameters.  The standard deviation of the sample of fitted
intervals for an element gives us an idea of how well the different
abundances agree internally with each other. A small value of $\sigma$
indicates that all intervals lead to similar abundances, which is a
good thing in our case. By examining the standard deviations obtained
when the temperature of the models is changed, we found the results to
be in better agreement when the effective temperature of our models is
near 82,000 K, which is the assumed one. As for the standard deviations
obtained with various log $g$ values, they do not point toward a favored
value of the surface gravity. Instead, we have to look at a comparison
of observed and synthetic spectra over a wavelength range featuring iron
lines originating from three ionization stages, namely
Fe~\textsc{v},~\textsc{vi}, and \textsc{vii} to have an indication of
the better value for the surface gravity (see Figure 7). This figure
suggests the assumed value of 6.2 dex to be the one giving the best
simultaneous match of the iron lines over the selected range. The
differences between our model spectra at different log $g$ are not as
striking as the one we obtained with different \teff~(see Figure 6), but
they are nevertheless significant. 

We also exploited the availability of an $Hipparcos$ parallax
measurement for \Bd~and compared the inferred distance with
spectroscopic distances estimated from several model spectra.
In a first step, we were able to derive an accurate determination of the
reddening between Earth and \Bd, $E(B-V) = 0.042 \pm 0.003$, thanks to
the high-precision optical photometry of \citet{land07}. Several
spectroscopic distances (with and without reddening correction) were
derived as indicated in Table 5. A comparison with the parallax distance
implies a relatively large value of the surface gravity and/or a small
mass for \Bd. We can reconcile our spectroscopic constraints with the
available parallax measurement only if the mass of \Bd~is significantly
less than the canonical value of 0.5 \msun~for a representative post-EHB
star. Assuming that the $Hipparcos$ measurement for \Bd~is fully
reliable and that our model atmospheres are reasonably realistic, we
must conclude that \Bd~is likely less massive than could have been
expected on the basis of standard evolutionary tracks.

Our analysis has allowed us to get a good idea of the atmospheric chemical
composition of \Bd. Its main constituents, in terms of atomic species,
have been analyzed and we were able to get abundances for eleven elements,
including the ones that influence the most the thermodynamical structure,
i.e., carbon, nitrogen, oxygen, and iron. This now allows us to
compute more realistic NLTE model atmospheres including the appropriate line
blanketing for \Bd. The value of 82,000 K for the effective temperature
of the star now seems quite robust and realistic uncertainties are
likely less than $\pm$5000 K. The case of the surface gravity is
somewhat more difficult, but we estimate conservatively that log $g$ = 
6.2$_{-0.1}^{+0.3}$ for \Bd. These informations should be a good
starting point for the study of the optical spectrum of the star. This
will follow in an upcoming paper.

\acknowledgments

This work was supported in part by the NSERC of Canada. G.F. also
acknowledges the contribution of the Canada Research Chair Program.
M.L. wishes to thank Patrick Dufour for help with the fitting routines.

\bibliographystyle{apj}
\bibliography{reference}

\newcommand{\SortNoop}[1]{}
\begin{thebibliography}{46}
\expandafter\ifx\csname natexlab\endcsname\relax\def\natexlab#1{#1}\fi

\bibitem[{{Asplund} {et~al.}(2009){Asplund}, {Grevesse}, {Sauval}, \&
  {Scott}}]{asp09}
{Asplund}, M., {Grevesse}, N., {Sauval}, A.~J., \& {Scott}, P. 2009, \araa, 47,
  481

\bibitem[{{Ayres}(2010)}]{ayres10}
{Ayres}, T.~R. 2010, \apjs, 187, 149

\bibitem[{{Bergeron} {et~al.}(1992){Bergeron}, {Saffer}, \& {Liebert}}]{ber92}
{Bergeron}, P., {Saffer}, R.~A., \& {Liebert}, J. 1992, \apj, 394, 228

\bibitem[{{Bruhweiler} \& {Dean}(1983)}]{bru83}
{Bruhweiler}, F.~C. \& {Dean}, C.~A. 1983, \apjl, 274, L87

\bibitem[{{Chayer} {et~al.}(2006){Chayer}, {Fontaine}, {Fontaine}, {Wesemael},
  \& {Dupuis}}]{chayer06}
{Chayer}, P., {Fontaine}, M., {Fontaine}, G., {Wesemael}, F., \& {Dupuis}, J.
  2006, Baltic Astronomy, 15, 131

\bibitem[{{Chayer} {et~al.}(1994){Chayer}, {LeBlanc}, {Fontaine}, {Wesemael},
  \& {Vennes}}]{cha94}
{Chayer}, P., {LeBlanc}, F., {Fontaine}, G., {Wesemael}, F., \& {Vennes}, S.
  1994, \apjl, 436, L161

\bibitem[{{Chayer} {et~al.}(2005){Chayer}, {Vennes}, {Dupuis}, \&
  {Kruk}}]{chayer05}
{Chayer}, P., {Vennes}, S., {Dupuis}, J., \& {Kruk}, J.~W. 2005, \apjl, 630,
  L169

\bibitem[{{Dixon} {et~al.}(2007){Dixon}, {Sahnow}, {Barrett}, {Civeit},
  {Dupuis}, {Fullerton}, {Godard}, {Hsu}, {Kaiser}, {Kruk}, {Lacour},
  {Lindler}, {Massa}, {Robinson}, {Romelfanger}, \& {Sonnentrucker}}]{dixon07}
{Dixon}, W.~V., {Sahnow}, D.~J., {Barrett}, P.~E., {Civeit}, T., {Dupuis}, J.,
  {Fullerton}, A.~W., {Godard}, B., {Hsu}, J.-C., {Kaiser}, M.~E., {Kruk},
  J.~W., {Lacour}, S., {Lindler}, D.~J., {Massa}, D., {Robinson}, R.~D.,
  {Romelfanger}, M.~L., \& {Sonnentrucker}, P. 2007, \pasp, 119, 527

\bibitem[{{Dorman} {et~al.}(1993){Dorman}, {Rood}, \& {O'Connell}}]{dor93}
{Dorman}, B., {Rood}, R.~T., \& {O'Connell}, R.~W. 1993, \apj, 419, 596

\bibitem[{{Dreizler} \& {Werner}(1993)}]{dre93}
{Dreizler}, S. \& {Werner}, K. 1993, \aap, 278, 199

\bibitem[{{Driebe} {et~al.}(1998){Driebe}, {Schoenberner}, {Bloecker}, \&
  {Herwig}}]{drie98}
{Driebe}, T., {Schoenberner}, D., {Bloecker}, T., \& {Herwig}, F. 1998, \aap,
  339, 123

\bibitem[{{Fontaine} {et~al.}(2012){Fontaine}, {Brassard}, {Charpinet}, \& {et
  al.}}]{fon12}
{Fontaine}, G., {Brassard}, P., {Charpinet}, S., \& {et al.} 2012, \aap, 539,
  A12

\bibitem[{{For} {et~al.}(2010){For}, {Green}, {Fontaine}, \& {et al.}}]{for10}
{For}, B.-Q., {Green}, E.~M., {Fontaine}, G., \& {et al.} 2010, \apj, 708, 253

\bibitem[{{Grevesse} \& {Sauval}(1998)}]{gre98}
{Grevesse}, N. \& {Sauval}, A.~J. 1998, \ssr, 85, 161

\bibitem[{{Haas} {et~al.}(1996){Haas}, {Dreizler}, {Heber}, {Jeffery}, \&
  {Werner}}]{haas96}
{Haas}, S., {Dreizler}, S., {Heber}, U., {Jeffery}, S., \& {Werner}, K. 1996,
  \aap, 311, 669

\bibitem[{{Heber} {et~al.}(2005){Heber}, {Drechsel}, {Karl}, {{\O}stensen}, B.,
  \& {Koester}}]{heb05}
{Heber}, U., {Drechsel}, H., {Karl}, C., {{\O}stensen}, R., B., \& {Koester},
  D. 2005, in Astronomical Society of the Pacific Conference Series, Vol. 334,
  14th European Workshop on White Dwarfs, ed. D.~{Koester} \& S.~{Moehler}, 357

\bibitem[{{Herbig}(1999)}]{her99}
{Herbig}, G.~H. 1999, \pasp, 111, 1144

\bibitem[{{Holberg} \& {Bergeron}(2006)}]{hol06}
{Holberg}, J.~B. \& {Bergeron}, P. 2006, \aj, 132, 1221

\bibitem[{{Landolt} \& {Uomoto}(2007)}]{land07}
{Landolt}, A.~U. \& {Uomoto}, A.~K. 2007, \aj, 133, 768

\bibitem[{{Lanz} \& {Hubeny}(2003)}]{lanz03}
{Lanz}, T. \& {Hubeny}, I. 2003, \apjs, 146, 417

\bibitem[{{Lanz} \& {Hubeny}(2007)}]{lanz07}
---. 2007, \apjs, 169, 83

\bibitem[{{Lanz} {et~al.}(1996){Lanz}, {Hubeny}, \& {de Koter}}]{lanz96}
{Lanz}, T., {Hubeny}, I., \& {de Koter}, A. 1996, Physica Scripta Volume T, 65,
  144

\bibitem[{{Latour} {et~al.}(2011){Latour}, {Fontaine}, {Brassard}, {Green},
  {Chayer}, \& {Randall}}]{lat11}
{Latour}, M., {Fontaine}, G., {Brassard}, P., {Green}, E.~M., {Chayer}, P., \&
  {Randall}, S.~K. 2011, \apj, 733, 100

\bibitem[{{Massey} \& {Gronwall}(1990)}]{mas90}
{Massey}, P. \& {Gronwall}, C. 1990, \apj, 358, 344

\bibitem[{{Moos} {et~al.}(2000){Moos}, {Cash}, {Cowie}, {Davidsen}, {Dupree},
  {Feldman}, {Friedman}, {Green}, {Green}, {Gry}, {Hutchings}, {Jenkins},
  {Linsky}, {Malina}, {Michalitsianos}, {Savage}, {Shull}, {Siegmund}, {Snow},
  {Sonneborn}, {Vidal-Madjar}, {Willis}, {Woodgate}, {York}, {Ake},
  {Andersson}, {Andrews}, {Barkhouser}, {Bianchi}, {Blair}, {Brownsberger},
  {Cha}, {Chayer}, {Conard}, {Fullerton}, {Gaines}, {Grange}, {Gummin},
  {Hebrard}, {Kriss}, {Kruk}, {Mark}, {McCarthy}, {Morbey}, {Murowinski},
  {Murphy}, {Oegerle}, {Ohl}, {Oliveira}, {Osterman}, {Sahnow}, {Saisse},
  {Sembach}, {Weaver}, {Welsh}, {Wilkinson}, \& {Zheng}}]{moos00}
{Moos}, H.~W., {Cash}, W.~C., {Cowie}, L.~L., {Davidsen}, A.~F., {Dupree},
  A.~K., {Feldman}, P.~D., {Friedman}, S.~D., {Green}, J.~C., {Green}, R.~F.,
  {Gry}, C., {Hutchings}, J.~B., {Jenkins}, E.~B., {Linsky}, J.~L., {Malina},
  R.~F., {Michalitsianos}, A.~G., {Savage}, B.~D., {Shull}, J.~M., {Siegmund},
  O.~H.~W., {Snow}, T.~P., {Sonneborn}, G., {Vidal-Madjar}, A., {Willis},
  A.~J., {Woodgate}, B.~E., {York}, D.~G., {Ake}, T.~B., {Andersson}, B.-G.,
  {Andrews}, J.~P., {Barkhouser}, R.~H., {Bianchi}, L., {Blair}, W.~P.,
  {Brownsberger}, K.~R., {Cha}, A.~N., {Chayer}, P., {Conard}, S.~J.,
  {Fullerton}, A.~W., {Gaines}, G.~A., {Grange}, R., {Gummin}, M.~A.,
  {Hebrard}, G., {Kriss}, G.~A., {Kruk}, J.~W., {Mark}, D., {McCarthy}, D.~K.,
  {Morbey}, C.~L., {Murowinski}, R., {Murphy}, E.~M., {Oegerle}, W.~R., {Ohl},
  R.~G., {Oliveira}, C., {Osterman}, S.~N., {Sahnow}, D.~J., {Saisse}, M.,
  {Sembach}, K.~R., {Weaver}, H.~A., {Welsh}, B.~Y., {Wilkinson}, E., \&
  {Zheng}, W. 2000, \apjl, 538, L1

\bibitem[{{Napiwotzki}(1993)}]{nap93}
{Napiwotzki}, R. 1993, Acta Astronomica, 43, 343

\bibitem[{{Napiwotzki}(1999)}]{nap99}
---. 1999, \aap, 350, 101

\bibitem[{{O'Toole} \& {Heber}(2006)}]{otoole06}
{O'Toole}, S.~J. \& {Heber}, U. 2006, \aap, 452, 579

\bibitem[{{Ramspeck} {et~al.}(2003){Ramspeck}, {Haas}, {Napiwotzki}, {Heber},
  {Deetjen}, \& {Dreizler}}]{ram03}
{Ramspeck}, M., {Haas}, S., {Napiwotzki}, R., {Heber}, U., {Deetjen}, J., \&
  {Dreizler}, S. 2003, in Astronomical Society of the Pacific Conference
  Series, Vol. 288, Stellar Atmosphere Modeling, ed. I.~{Hubeny}, D.~{Mihalas},
  \& K.~{Werner}, 161

\bibitem[{{Randall} {et~al.}(2007){Randall}, {Green}, {Van Grootel},
  {Fontaine}, {Charpinet}, {Lesser}, {Brassard}, {Sugimoto}, {Chayer}, {Fay},
  {Wroblewski}, {Daniel}, {Story}, \& {Fitzgerald}}]{ran07}
{Randall}, S.~K., {Green}, E.~M., {Van Grootel}, V., {Fontaine}, G.,
  {Charpinet}, S., {Lesser}, M., {Brassard}, P., {Sugimoto}, T., {Chayer}, P.,
  {Fay}, A., {Wroblewski}, P., {Daniel}, M., {Story}, S., \& {Fitzgerald}, T.
  2007, \aap, 476, 1317

\bibitem[{{Rauch} {et~al.}(2007){Rauch}, {Ziegler}, {Werner}, {Kruk},
  {Oliveira}, {Vande Putte}, {Mignani}, \& {Kerber}}]{rauch07}
{Rauch}, T., {Ziegler}, M., {Werner}, K., {Kruk}, J.~W., {Oliveira}, C.~M.,
  {Vande Putte}, D., {Mignani}, R.~P., \& {Kerber}, F. 2007, \aap, 470, 317

\bibitem[{{Saffer} {et~al.}(1994){Saffer}, {Bergeron}, {Koester}, \&
  {Liebert}}]{saf94}
{Saffer}, R.~A., {Bergeron}, P., {Koester}, D., \& {Liebert}, J. 1994, \apj,
  432, 351

\bibitem[{{Sahnow} {et~al.}(2000){Sahnow}, {Moos}, {Ake}, {Andersen},
  {Andersson}, {Andre}, {Artis}, {Berman}, {Blair}, {Brownsberger}, {Calvani},
  {Chayer}, {Conard}, {Feldman}, {Friedman}, {Fullerton}, {Gaines}, {Gawne},
  {Green}, {Gummin}, {Jennings}, {Joyce}, {Kaiser}, {Kruk}, {Lindler}, {Massa},
  {Murphy}, {Oegerle}, {Ohl}, {Roberts}, {Romelfanger}, {Roth}, {Sankrit},
  {Sembach}, {Shelton}, {Siegmund}, {Silva}, {Sonneborn}, {Vaclavik}, {Weaver},
  \& {Wilkinson}}]{sahnow00}
{Sahnow}, D.~J., {Moos}, H.~W., {Ake}, T.~B., {Andersen}, J., {Andersson},
  B.-G., {Andre}, M., {Artis}, D., {Berman}, A.~F., {Blair}, W.~P.,
  {Brownsberger}, K.~R., {Calvani}, H.~M., {Chayer}, P., {Conard}, S.~J.,
  {Feldman}, P.~D., {Friedman}, S.~D., {Fullerton}, A.~W., {Gaines}, G.~A.,
  {Gawne}, W.~C., {Green}, J.~C., {Gummin}, M.~A., {Jennings}, T.~B., {Joyce},
  J.~B., {Kaiser}, M.~E., {Kruk}, J.~W., {Lindler}, D.~J., {Massa}, D.,
  {Murphy}, E.~M., {Oegerle}, W.~R., {Ohl}, R.~G., {Roberts}, B.~A.,
  {Romelfanger}, M.~L., {Roth}, K.~C., {Sankrit}, R., {Sembach}, K.~R.,
  {Shelton}, R.~L., {Siegmund}, O.~H.~W., {Silva}, C.~J., {Sonneborn}, G.,
  {Vaclavik}, S.~R., {Weaver}, H.~A., \& {Wilkinson}, E. 2000, \apjl, 538, L7

\bibitem[{{Schlegel} {et~al.}(1998){Schlegel}, {Finkbeiner}, \&
  {Davis}}]{dust98}
{Schlegel}, D.~J., {Finkbeiner}, D.~P., \& {Davis}, M. 1998, \apj, 500, 525

\bibitem[{{Schoenberner}(1983)}]{scho83}
{Schoenberner}, D. 1983, \apj, 272, 708

\bibitem[{{Seaton}(1979)}]{sea79}
{Seaton}, M.~J. 1979, \mnras, 187, 73P

\bibitem[{{Sonneborn} {et~al.}(2002){Sonneborn}, {Andr{\'e}}, {Oliveira},
  {H{\'e}brard}, {Howk}, {Tripp}, {Chayer}, {Friedman}, {Kruk}, {Jenkins},
  {Lemoine}, {Moos}, {Oegerle}, {Sembach}, \& {Vidal-Madjar}}]{son02}
{Sonneborn}, G., {Andr{\'e}}, M., {Oliveira}, C., {H{\'e}brard}, G., {Howk},
  J.~C., {Tripp}, T.~M., {Chayer}, P., {Friedman}, S.~D., {Kruk}, J.~W.,
  {Jenkins}, E.~B., {Lemoine}, M., {Moos}, H.~W., {Oegerle}, W.~R., {Sembach},
  K.~R., \& {Vidal-Madjar}, A. 2002, \apjs, 140, 51

\bibitem[{{Stroeer} {et~al.}(2007){Stroeer}, {Heber}, {Lisker}, {Napiwotzki},
  {Dreizler}, {Christlieb}, \& {Reimers}}]{stro07}
{Stroeer}, A., {Heber}, U., {Lisker}, T., {Napiwotzki}, R., {Dreizler}, S.,
  {Christlieb}, N., \& {Reimers}, D. 2007, \aap, 462, 269

\bibitem[{{van Leeuwen}(2007)}]{hip07}
{van Leeuwen}, F. 2007, \aap, 474, 653

\bibitem[{{Vennes} {et~al.}(2005){Vennes}, {Chayer}, \& {Dupuis}}]{vennes05}
{Vennes}, S., {Chayer}, P., \& {Dupuis}, J. 2005, \apjl, 622, L121

\bibitem[{{Wassermann} {et~al.}(2010){Wassermann}, {Werner}, {Rauch}, \&
  {Kruk}}]{wass10}
{Wassermann}, D., {Werner}, K., {Rauch}, T., \& {Kruk}, J.~W. 2010, \aap, 524,
  A9

\bibitem[{{Werner}(1996)}]{wer96}
{Werner}, K. 1996, \apjl, 457, L39

\bibitem[{{Werner} \& {Dreizler}(1994)}]{wer94ni}
{Werner}, K. \& {Dreizler}, S. 1994, \aap, 286, L31

\bibitem[{{Werner} {et~al.}(1994){Werner}, {Heber}, \& {Fleming}}]{wer94}
{Werner}, K., {Heber}, U., \& {Fleming}, T. 1994, \aap, 284, 907

\bibitem[{{Werner} {et~al.}(2005){Werner}, {Rauch}, \& {Kruk}}]{wer05}
{Werner}, K., {Rauch}, T., \& {Kruk}, J.~W. 2005, \aap, 433, 641

\bibitem[{{Werner} {et~al.}(2012){Werner}, {Rauch}, {Ringat}, \&
  {Kruk}}]{werner12}
{Werner}, K., {Rauch}, T., {Ringat}, E., \& {Kruk}, J.~W. 2012, \apjl, 753, L7

\end{thebibliography}

\vspace*{1\baselineskip}
\centerline{\bf{ONLINE FIGURES}}

\noindent Online Figure -- The online figure present the totality of our
fitted spectral intervals for all the atomic species. The fits were done with
models having \teff~= 82,000 K, log $g$ = 6.2, and log \nhe~=$-$1.0. The
red curve shows the result of the fitting procedure for the element
mentionned on each panel and the abundance obtained is expressed as log
$N$(X)/$N$(H). a) The fit showed in the lower left panel was not
included in our mean abundance for nitrogen. c) The O~\textsc{iv} line
at 1338.6 \AA~ is fitted twice because each line comes from two
different STIS orders. They were both included in the mean
abundance. The lower right panel shows a view of He~\textsc{ii}
$\lambda$1640, with the oxygen abundance fixed to its determined
value. g) In the middle top panel, the line at the left is
Co~\textsc{vi} $\lambda$1226.4, an element not included in our model
atmospheres.

\begin{figure*}[t!]
\begin{center}
\includegraphics[scale=.60,angle=270]{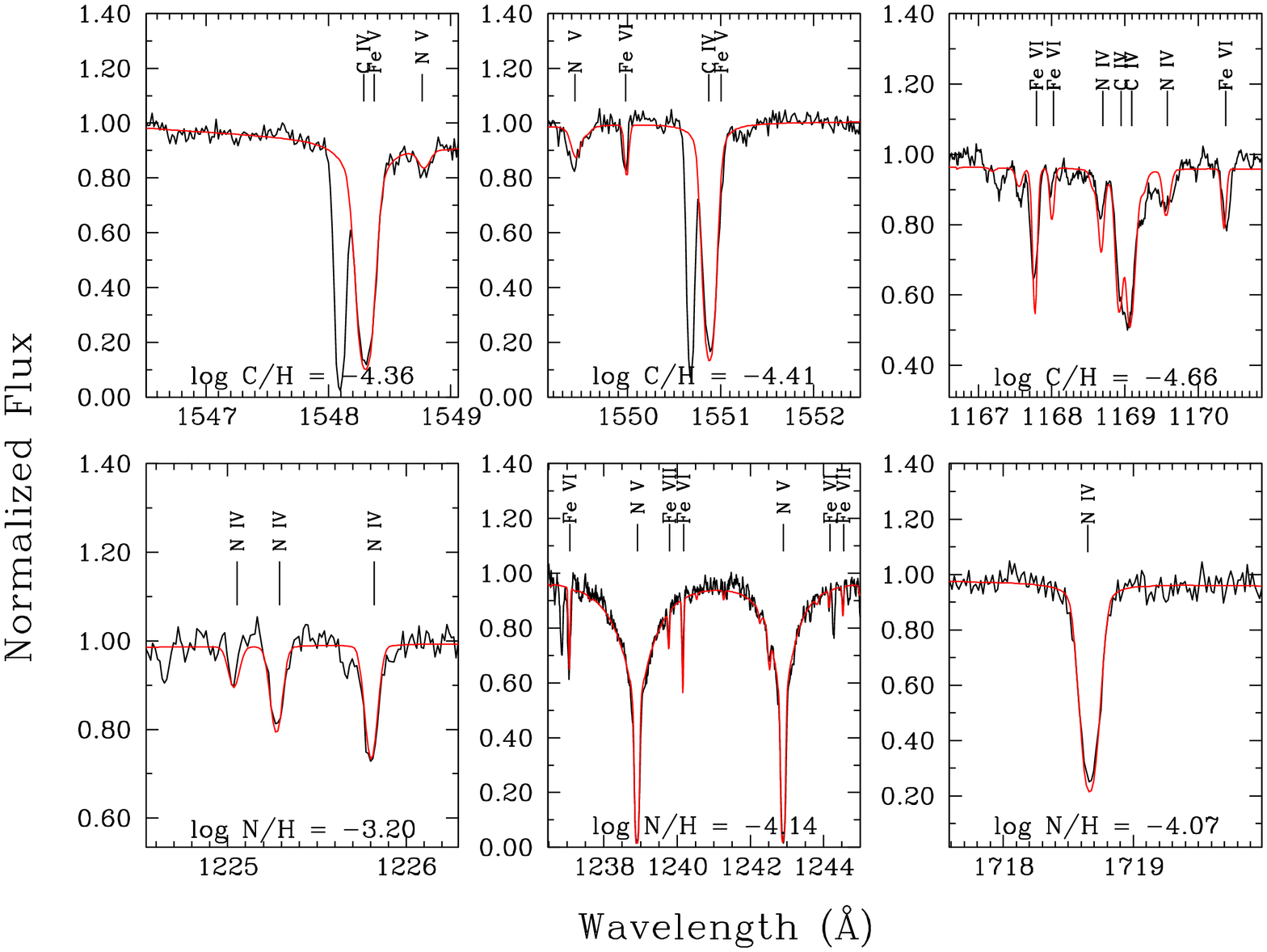}
\end{center}
\begin{flushright}
Online Figure panel a
\end{flushright}
\begin{center}
\includegraphics[scale=.60,angle=270]{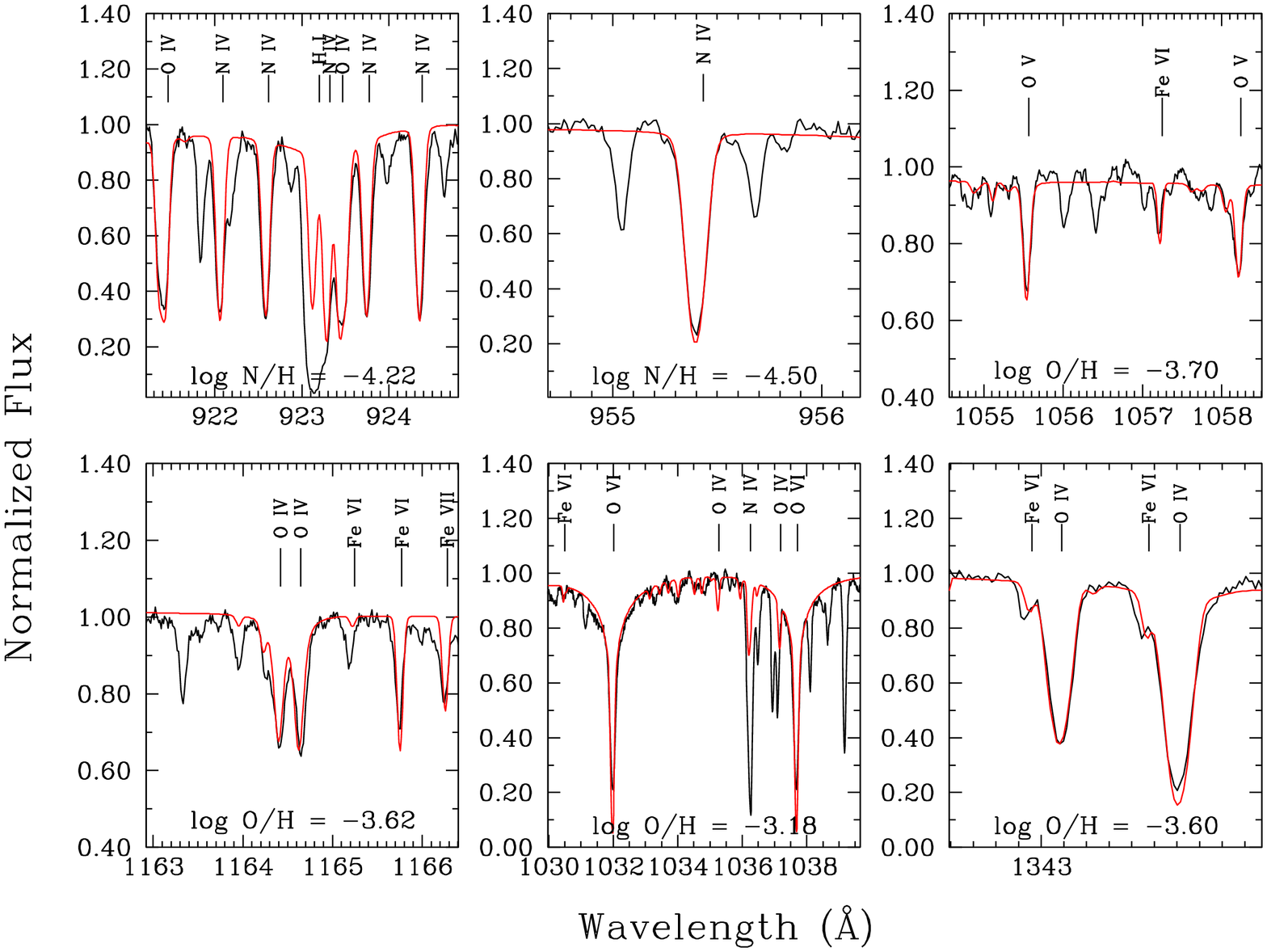}
\end{center}
\begin{flushright}
Online Figure panel b
\end{flushright}
\end{figure*}

\begin{figure*}[t!]
\begin{center}
\includegraphics[scale=.60,angle=270]{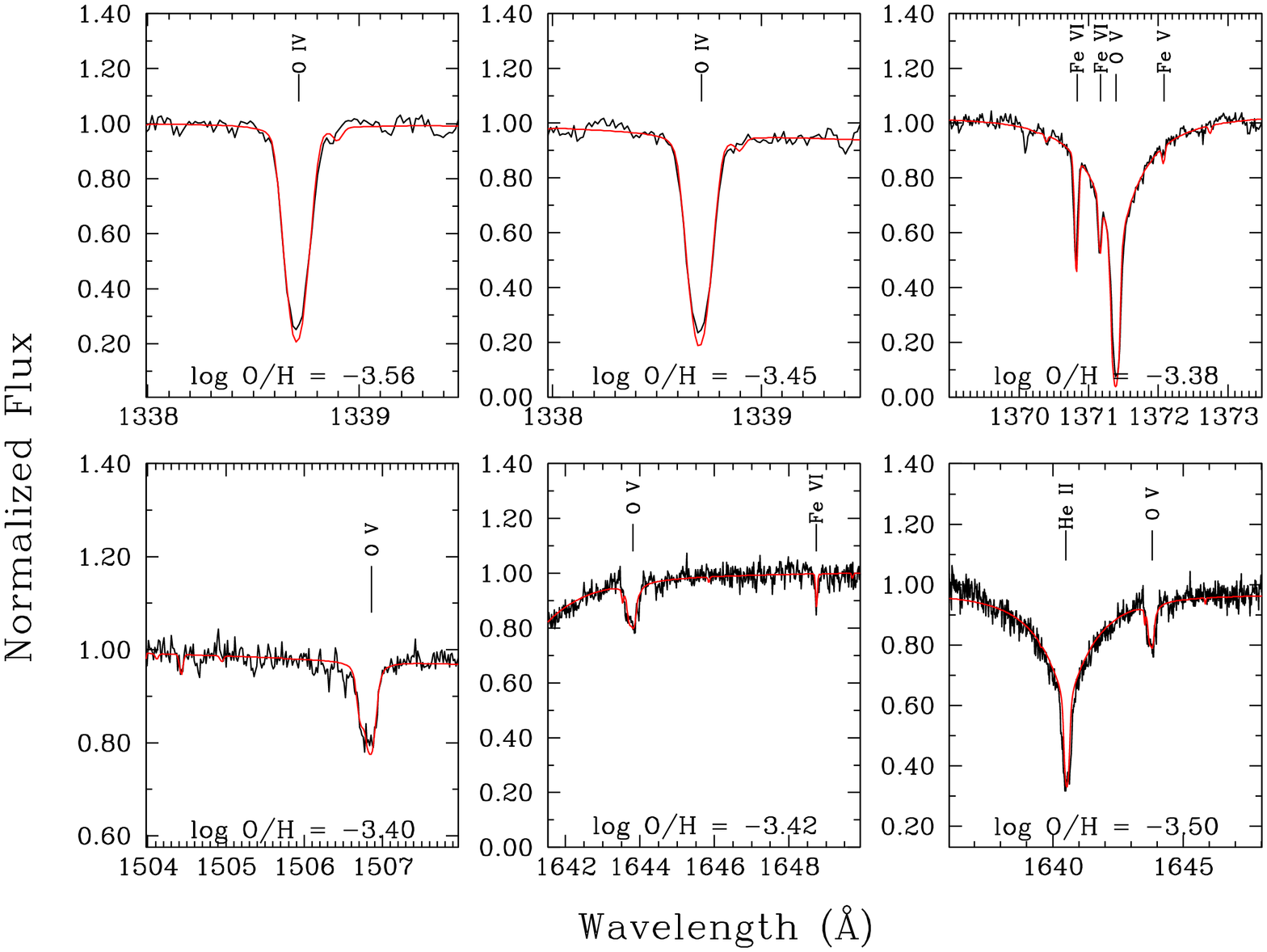}
\end{center}
\begin{flushright}
Online Figure panel c
\end{flushright}
\begin{center}
\includegraphics[scale=.60,angle=270]{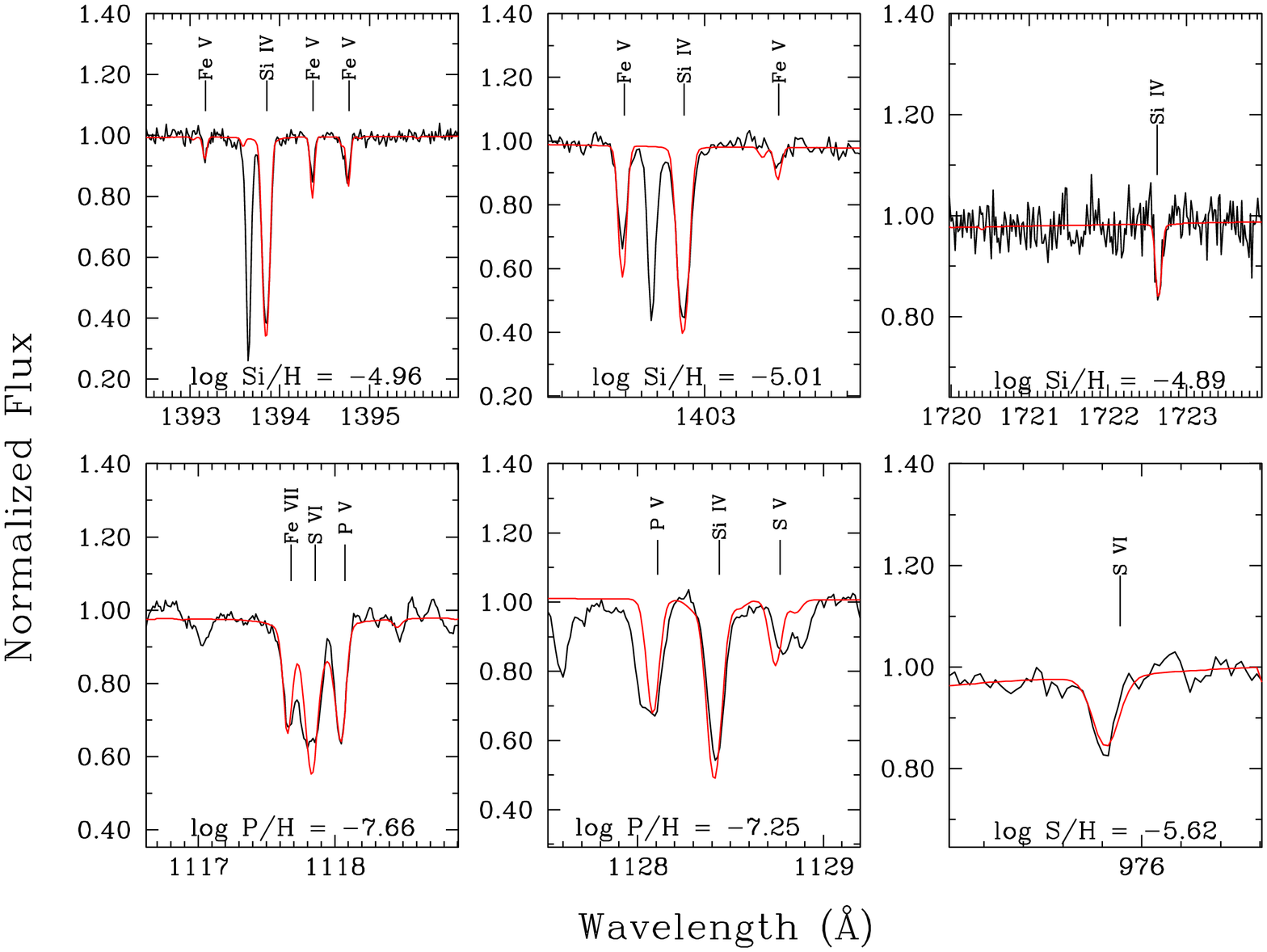}
\end{center}
\begin{flushright}
Online Figure panel d
\end{flushright}
\end{figure*}

\begin{figure*}[t!]
\begin{center}
\includegraphics[scale=.60,angle=270]{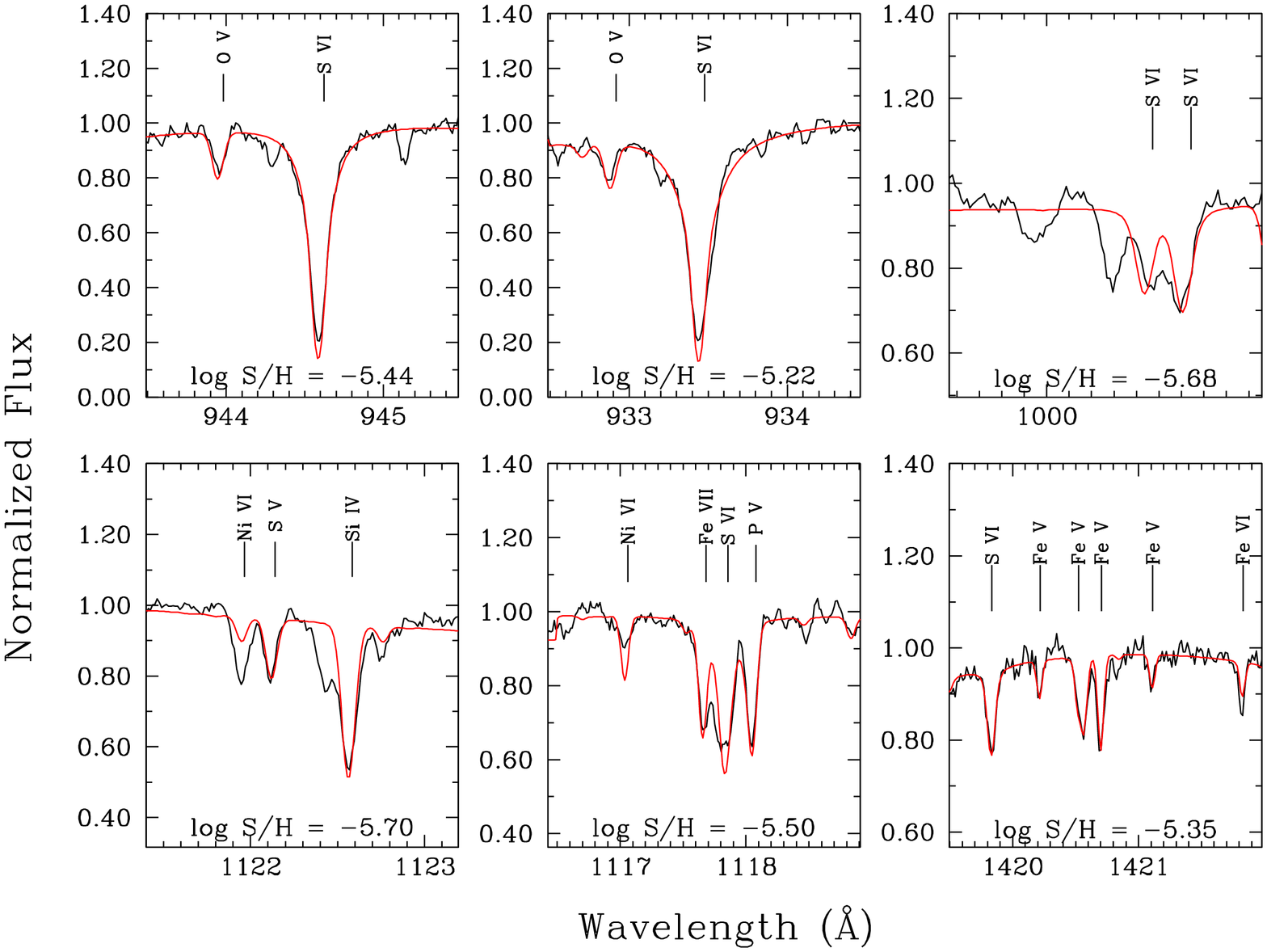}
\end{center}
\begin{flushright}
Online Figure panel e
\end{flushright}
\begin{center}
\includegraphics[scale=.60,angle=270]{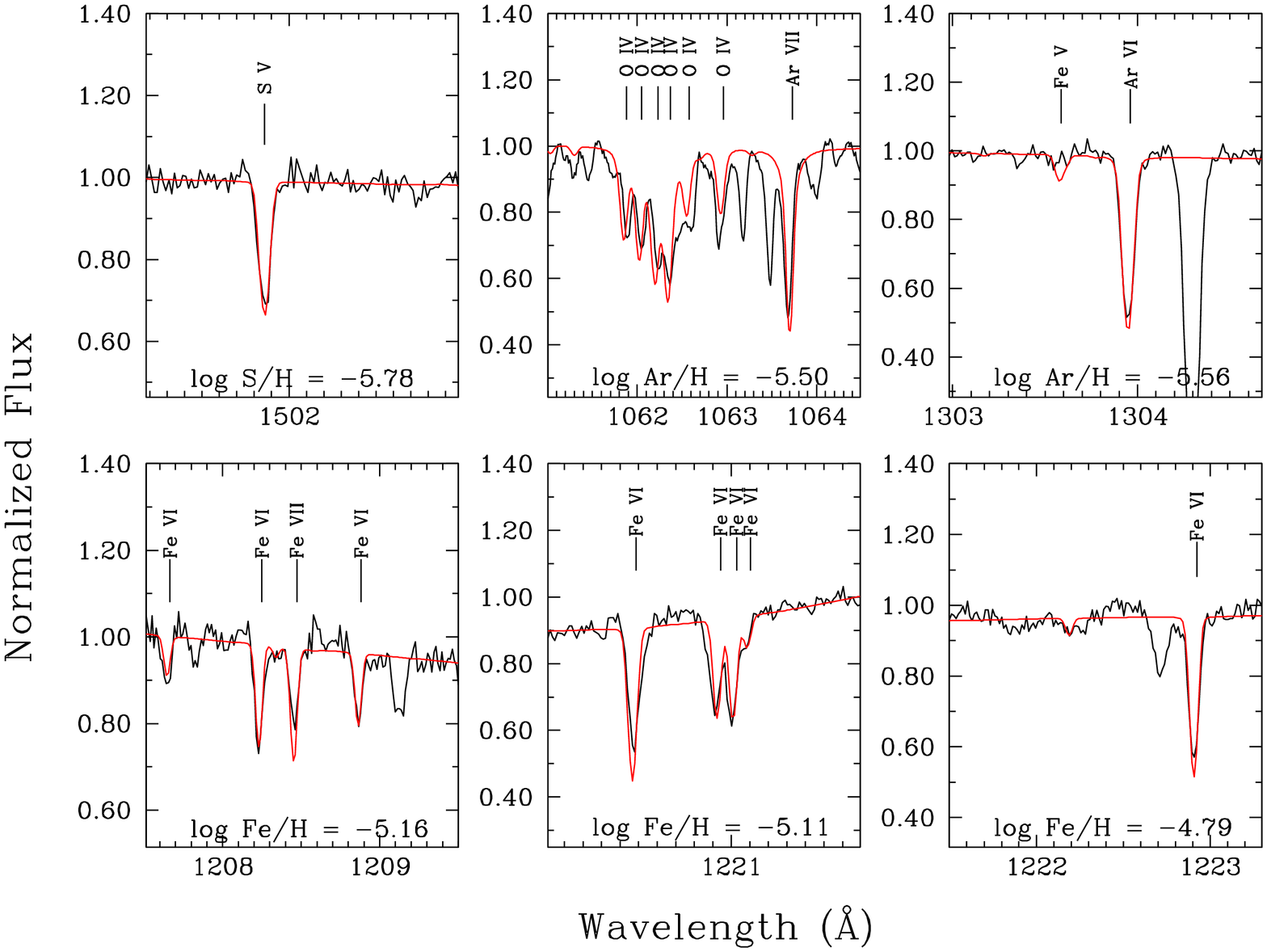}
\end{center}
\begin{flushright}
Online Figure panel f
\end{flushright}
\end{figure*}

\begin{figure*}[t!]
\begin{center}
\includegraphics[scale=.60,angle=270]{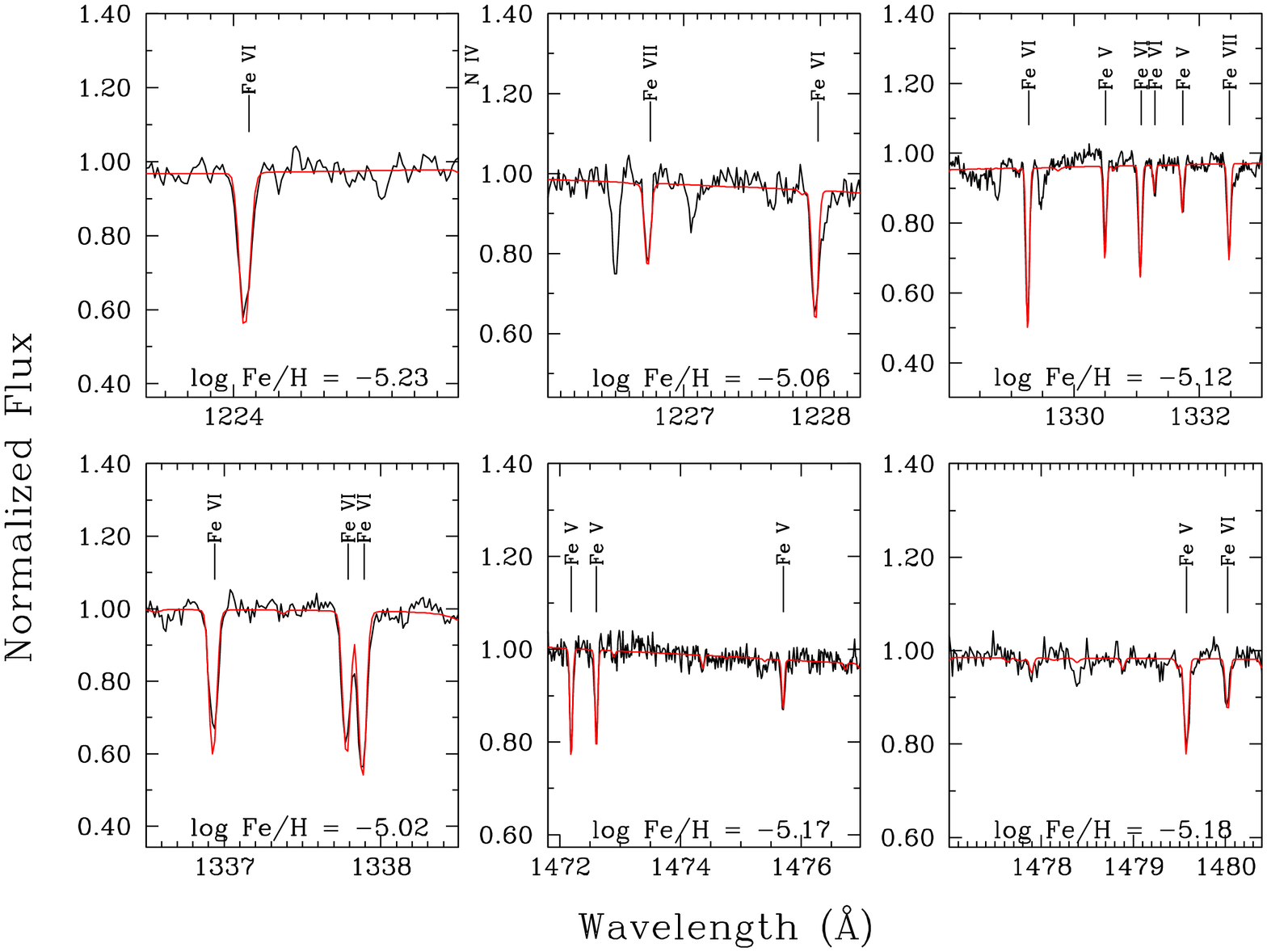}
\end{center}
\begin{flushright}
Online Figure panel g
\end{flushright}
\begin{center}
\includegraphics[scale=.60,angle=270]{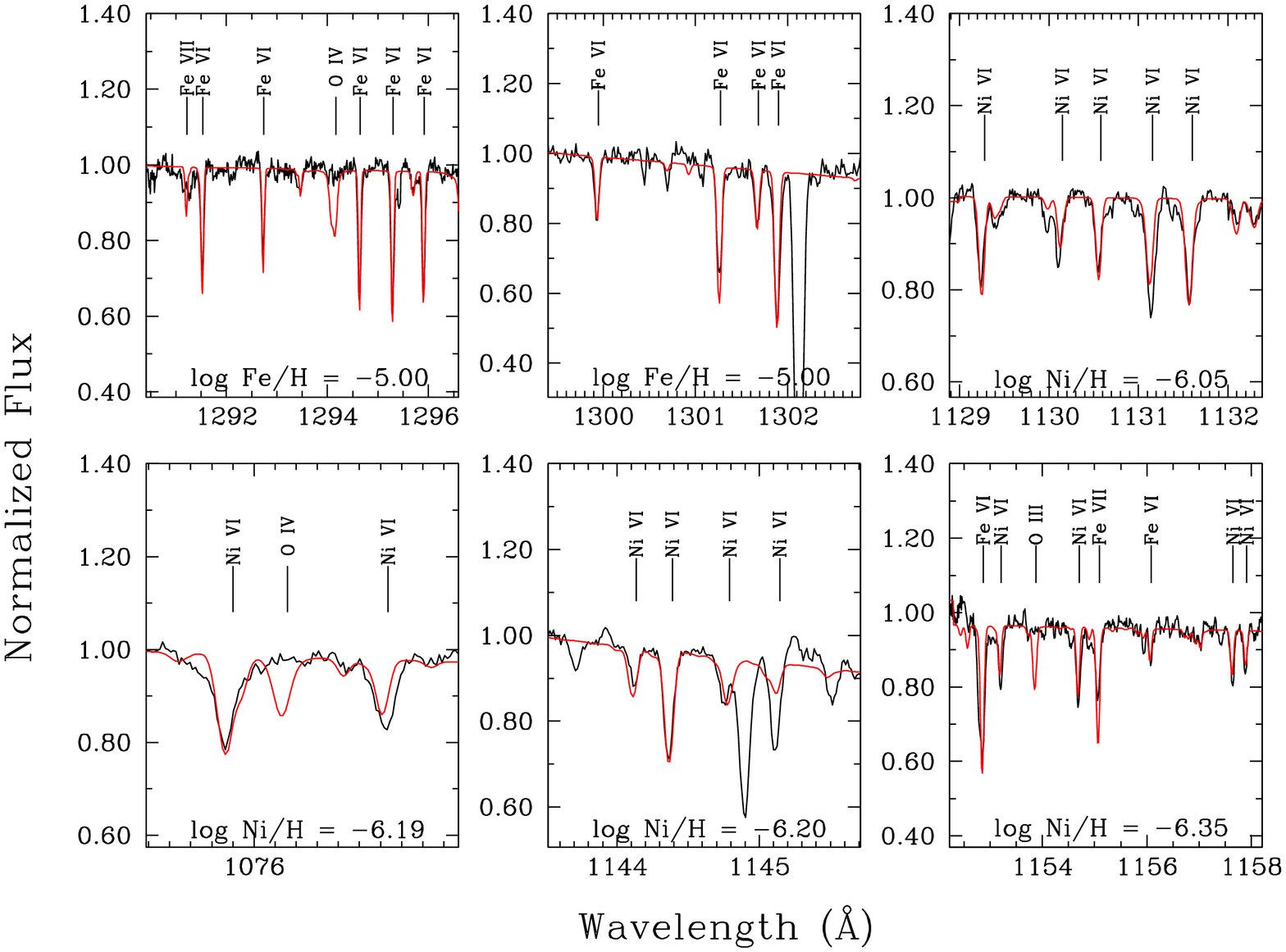}
\end{center}
\begin{flushright}
Online Figure panel h
\end{flushright}
\end{figure*}

\begin{figure*}[t!]
\begin{center}
\includegraphics[scale=.60,angle=270]{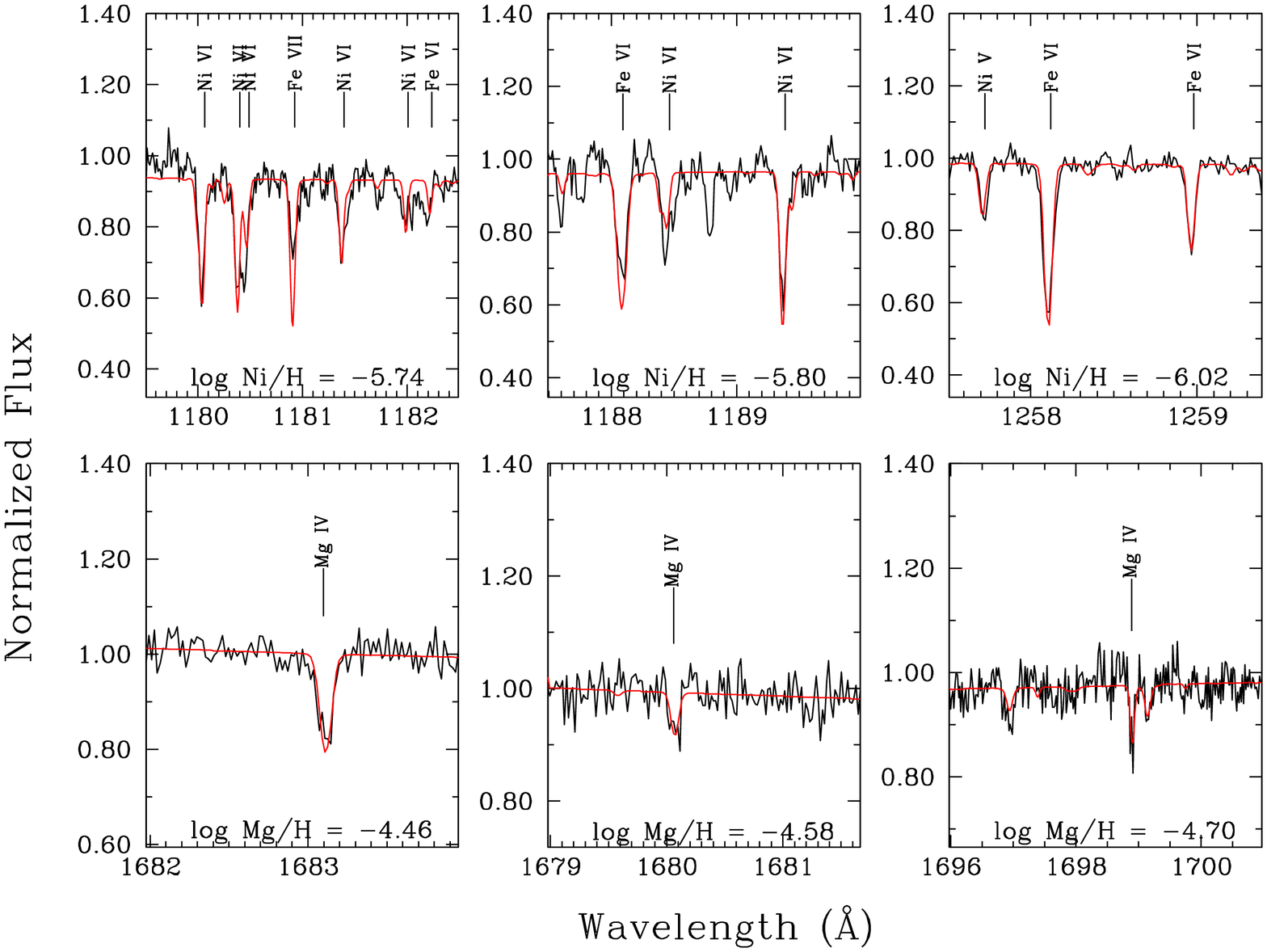}
\end{center}
\begin{flushright}
Online Figure panel i
\end{flushright}
\begin{center}
\includegraphics[scale=.60,angle=270]{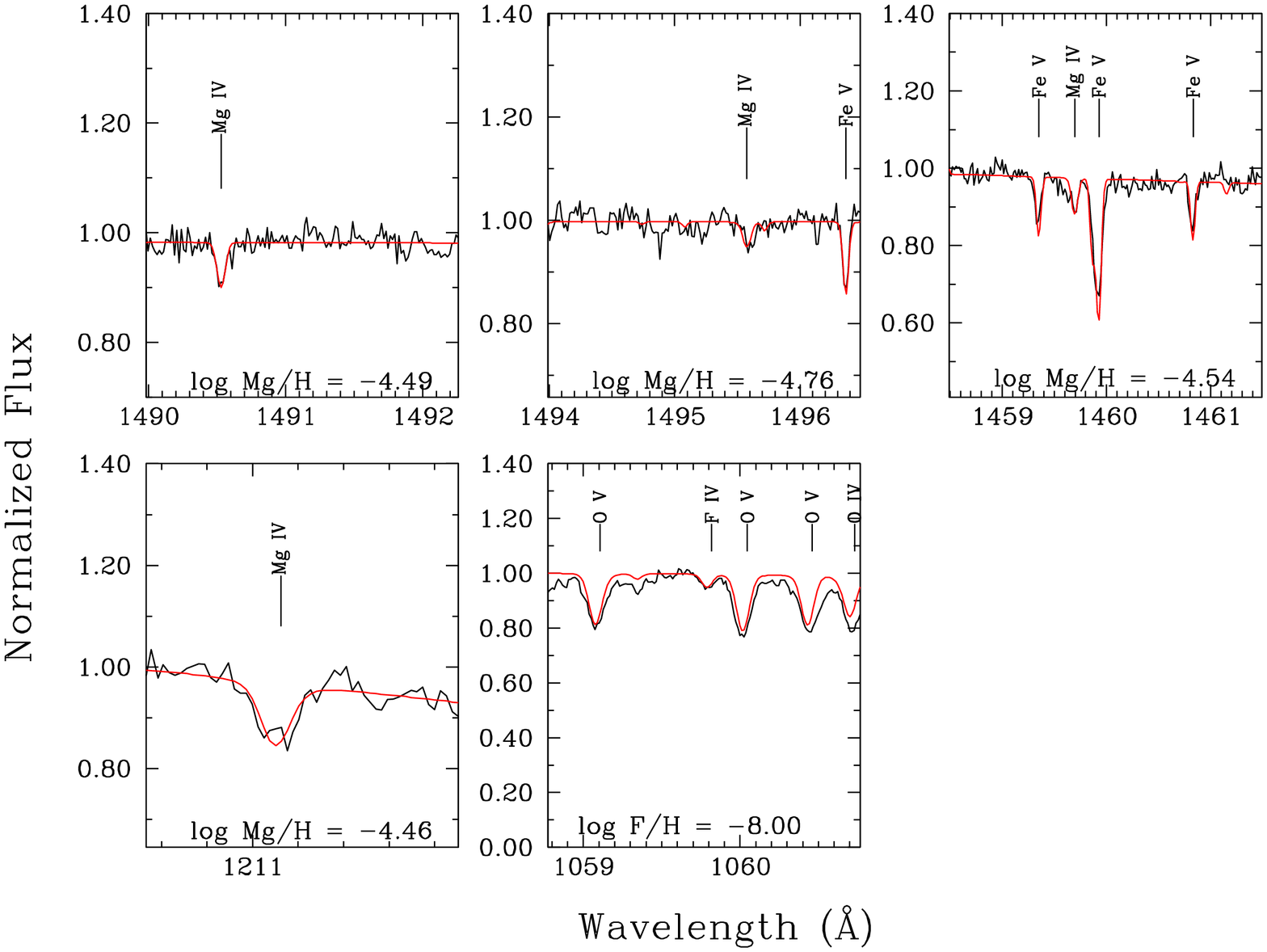}
\end{center}
\begin{flushright}
Online Figure panel j
\end{flushright}
\end{figure*}

\end{document}